\documentclass[twocolumn]{aastex631}

\begin{document}

\vspace*{-0.50cm}
\title{Magellanic System Stars Identified in SMACS\,J0723.3-7327 James Webb Space Telescope Early Release Observations Images} 

\author[0000-0002-7265-7920]{Jake Summers} 
\affiliation{School of Earth and Space Exploration, Arizona State University,
Tempe, AZ 85287-1404, USA}

\author[0000-0001-8156-6281]{Rogier A. Windhorst}
\affiliation{School of Earth and Space Exploration, Arizona State University,
Tempe, AZ 85287-1404, USA}

\author[0000-0003-3329-1337]{Seth H. Cohen} 
\affiliation{School of Earth and Space Exploration, Arizona State University,
Tempe, AZ 85287-1404, USA}

\author[0000-0003-1268-5230]{Rolf A. Jansen} 
\affiliation{School of Earth and Space Exploration, Arizona State University,
Tempe, AZ 85287-1404, USA}

\author[0000-0001-6650-2853]{Timothy Carleton} 
\affiliation{School of Earth and Space Exploration, Arizona State University,
Tempe, AZ 85287-1404, USA}

\author[0000-0001-9394-6732]{Patrick S. Kamieneski} 
\affiliation{School of Earth and Space Exploration, Arizona State University, Tempe, AZ 85287-1404, USA}

\author[0000-0002-4884-6756]{Benne W. Holwerda} 
\affiliation{Department of Physics and Astronomy, University of Louisville,
Louisville KY 40292, USA} 

\author[0000-0003-1949-7638]{Christopher J. Conselice} 
\affiliation{Jodrell Bank Centre for Astrophysics, Alan Turing Building, 
University of Manchester, Oxford Road, Manchester M13 9PL, UK}

\author[0000-0003-4875-6272]{Nathan J. Adams} 
\affiliation{Jodrell Bank Centre for Astrophysics, Alan Turing Building, 
University of Manchester, Oxford Road, Manchester M13 9PL, UK}

\author[0000-0003-1625-8009]{Brenda Frye} 
\affiliation{Steward Observatory, University of Arizona, 933 N Cherry Ave,
Tucson, AZ, 85721-0009, USA}

\author[0000-0001-9065-3926]{Jose M. Diego} 
\affiliation{Instituto de F\'isica de Cantabria (CSIC-UC). Avenida. Los Castros
s/n. 39005 Santander, Spain}

\author[0000-0001-9262-9997]{Christopher N. A. Willmer} 
\affiliation{Steward Observatory, University of Arizona, 933 N Cherry Ave,
Tucson, AZ, 85721-0009, USA}

\author[0000-0002-6150-833X]{Rafael Ortiz III}
\affiliation{School of Earth and Space Exploration, Arizona State University,
Tempe, AZ 85287-1404, USA}

\author[0000-0003-0202-0534]{Cheng Cheng} 
\affiliation{Chinese Academy of Sciences, National Astronomical Observatories,
CAS, Beijing 100101, China}

\author[0000-0001-9369-6921]{Alex Pigarelli} 
\affiliation{School of Earth and Space Exploration, Arizona State University,
Tempe, AZ 85287-1404, USA}

\author[0000-0003-0429-3579]{Aaron Robotham} 
\affiliation{International Centre for Radio Astronomy Research (ICRAR) and the
International Space Centre (ISC), The University of Western Australia, M468,
35 Stirling Highway, Crawley, WA 6009, Australia}

\author[0000-0002-9816-1931]{Jordan C. J. D'Silva} 
\affiliation{International Centre for Radio Astronomy Research (ICRAR) and the
International Space Centre (ISC), The University of Western Australia, M468,
35 Stirling Highway, Crawley, WA 6009, Australia}
\affiliation{ARC Centre of Excellence for All Sky Astrophysics in 3 Dimensions
(ASTRO 3D), Australia}

\author[0000-0001-9052-9837]{Scott Tompkins} 
\affiliation{School of Earth and Space Exploration, Arizona State University,
Tempe, AZ 85287-1404, USA}

\author[0000-0001-9491-7327]{Simon P. Driver} 
\affiliation{International Centre for Radio Astronomy Research (ICRAR) and the
International Space Centre (ISC), The University of Western Australia, M468,
35 Stirling Highway, Crawley, WA 6009, Australia}

\author[0000-0001-7592-7714]{Haojing Yan} 
\affiliation{Department of Physics and Astronomy, University of Missouri,
Columbia, MO 65211, USA}

\author[0000-0001-7410-7669]{Dan Coe} 
\affiliation{AURA for the European Space Agency (ESA), Space Telescope Science
Institute, 3700 San Martin Drive, Baltimore, MD 21210, USA}

\author[0000-0001-9440-8872]{Norman Grogin} 
\affiliation{Space Telescope Science Institute, 3700 San Martin Drive, 
Baltimore, MD 21210, USA}

\author[0000-0002-6610-2048]{Anton Koekemoer} 
\affiliation{Space Telescope Science Institute, 3700 San Martin Drive,
Baltimore, MD 21210, USA}

\author[0000-0001-6434-7845]{Madeline A. Marshall} 
\affiliation{National Research Council of Canada, Herzberg Astronomy \&
Astrophysics Research Centre, 5071 West Saanich Road, Victoria, BC V9E 2E7, 
Canada}
\affiliation{ARC Centre of Excellence for All Sky Astrophysics in 3 Dimensions
(ASTRO 3D), Australia}

\author[0000-0003-3382-5941]{Nor Pirzkal} 
\affiliation{Space Telescope Science Institute, 3700 San Martin Drive,
Baltimore, MD 21210, USA}

\author[0000-0003-0894-1588]{Russell E. Ryan, Jr.} 
\affiliation{Space Telescope Science Institute, 3700 San Martin Drive, 
Baltimore, MD 21210, USA}

\email{jssumme1@asu.edu}

\shorttitle{Magellanic Stars in front of SMACS\,0723}
\shortauthors{Summers et al.}

\begin{abstract} 
We identify 71 distant stars in JWST/NIRCam ERO images of the field of galaxy cluster SMACS\,J0723.3-7327 (SMACS\,0723). Given the relatively small ($\sim$$10^{\circ}$) angular separation between SMACS\,0723 and the Large Magellanic Cloud, it is likely that these stars are associated with the LMC outskirts or Leading Arm. This is further bolstered by a spectral energy distribution analysis, which suggests an excess of stars at a physical distance of $40-100$ kpc, consistent with being associated with or located behind the Magellanic system. In particular, we find that the overall surface density of stars brighter than 27.0 mag in the field of SMACS\,0723 is $\sim$2.3 times that of stars in a blank field with similar galactic latitude (the North Ecliptic Pole Time Domain Field), and that the density of stars in the SMACS\,0723 field with SED-derived distances consistent with the Magellanic system is $\sim$6.1 times larger than that of the blank field. The candidate stars at these distances are consistent with a stellar population at the same distance modulus with [Fe/H]\,$= -1.0$ and an age of $\sim$$5.0$ Gyr. On the assumption that all of the 71 stars are associated with the LMC, then the stellar density of the LMC at the location of the SMACS\,0723 field is $\sim$$740$ stars kpc$^{-3}$, which helps trace the density of stars in the LMC outskirts.
\end{abstract}


\section{Introduction} \label{sec1}

The Magellanic Clouds are the most massive satellite galaxies of the Milky Way \citep{mcconnachie+2012}, so they can provide a uniquely detailed look at how massive satellite galaxies and Milky Way-like host galaxies interact. Also, they are the only nearby satellites of the Milky Way that are not devoid of gas \citep{putnam21}. Those two seem connected---most satellite galaxies of this mass are star forming \citep{wheeler+14}, but how tidal and ram-pressure stripping processes affect galaxies like the LMC and SMC in orbit around the Milky Way is not fully understood. A more detailed view of the structure and dynamics of the Magellanic Clouds can help improve our understanding of the interaction between the Clouds and the Milky Way, and in turn these environmental processes. The detailed orbital history of the Magellanic Clouds can provide clues about the properties (mass and concentration) of the Milky Way (e.g., \citealt{santos-santos+21}).

The Magellanic system (See \citealt{lmc_overview} for a review) consists of the Large Magellanic Cloud (LMC), the Small Magellanic Cloud (SMC), the Bridge (between the LMC and SMC; \citealt{Bridge_discovery}), the Magellanic Stream (behind the LMC and SMC; \citealt{mathewson_discovery}), and the Leading Arm (ahead of the LMC and SMC; \citealt{putman1998}). The Leading Arm was first identified by \citet{putman1998} in the \ion{H}{1} Parkes All-Sky Survey as a thin region leading the LMC-SMC system at higher velocities. \citet{Lu_1998} further connected the Leading Arm to the LMC and SMC by finding similarities to LMC and SMC S/Fe abundance ratios in a high-velocity cloud in the Leading Arm.

As a key rung in the distance ladder \citep{riess_H0}, the stellar populations of the LMC and SMC have been studied in detail. The distance modulus to the LMC has been measured to be $18.48 \pm 0.05$ (based on a combination of Cepheids, Luminous Red Variables, RR Lyrae Stars, Red Clump Stars, and Eclipsing Binaries), which corresponds to a distance between $48.5 \ \text{kpc}$ and $50.8 \ \text{kpc}$ \citep{Walker_2011, pietrzynski19, riess19}. The SMC is around $10 \ \text{kpc}$ farther away, measured with similar techniques \citep{Graczyk_2013}. The velocity distribution of \ion{H}{1} clouds in the Leading Arm is consistent with a distance between $40$ and $70$ kpc \citep{venzmer_2012}. Some simulations of the Magellanic Stream predict that it is $\sim$$20 \ \text{kpc}$ from the Sun at its closest point, with a stellar component that is even closer \citep{Lucchini2021}.

The Leading Arm spans around $70^{\circ}$ in the sky, has an inclination against the sky of $\sim$$13.6^{\circ}$ (almost face-on), and spans around 52 kpc \citep{bruns_2005, venzmer_2012}. Most observations of the Leading Arm are of \ion{H}{1} gas, of which \citet{venzmer_2012} estimated a lower limit \ion{H}{1} mass of $3.8 \times 10^7 \ M_{\odot}$, compared to $4.4 \times 10^8 \ M_{\odot}$ for the LMC and $4.0 \times 10^8 \ M_{\odot}$ for the SMC \citep{bruns_2005}. However, the overall properties of the Leading Arm are not as well studied as the Clouds themselves.

\begin{figure}[t]
    \centerline{\fig{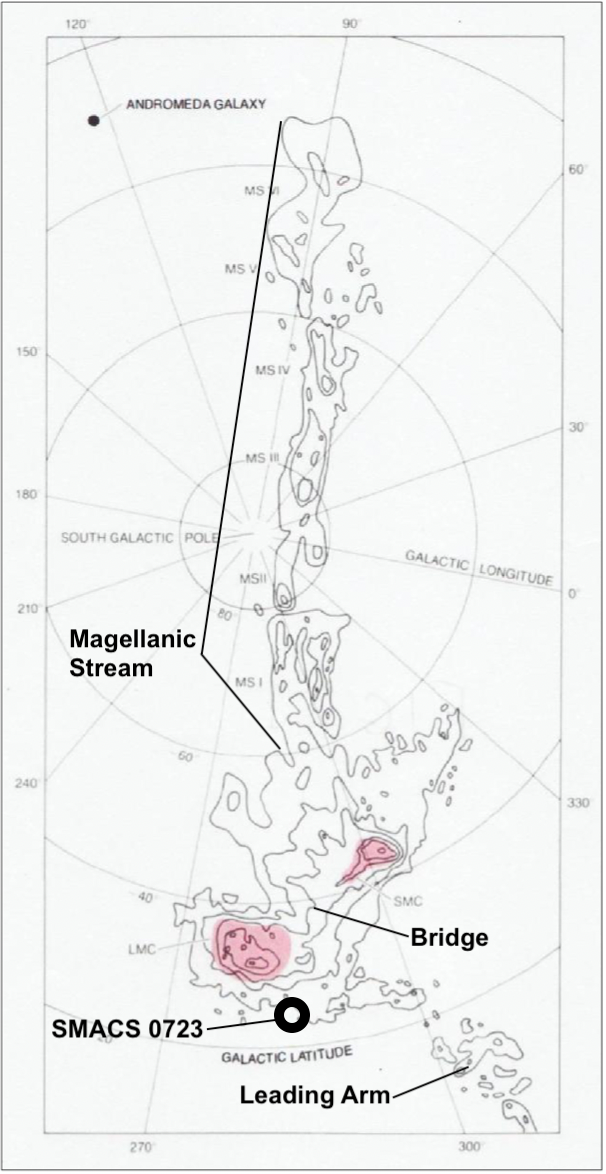}{0.5\textwidth}{}}
    \caption{Map of \ion{H}{1} in the LMC-SMC system, in galactic coordinates, showing the location of the field of SMACS\,0723. The SMACS\,0723 field is near the Leading Arm and is part of the LMC outskirts, only $\sim$$10\degr$ from the center of the LMC. This image is reproduced from Fig. 3 of \citet{mathewson_img}, with explicit permission from the editors of JAHH, with additional annotations showing the location of the Bridge, Leading Arm, Magellanic Stream, and SMACS\,0723 field. The outermost contours represent $10^{19}$ atoms cm$^{-2}$.}
    \label{fig:lmc_map}
\end{figure}

The formation of the Leading Arm must involve tidal forces, although additional forces such as ram pressure stripping may have been involved in its formation \citep{nidever_2008}. At the same time, the Magellanic Stream could have formed due to ram pressure stripping \citep{meurer_rp, moore_rp} or by blowout from supernovae \citep{nidever_2008}. With ram pressure forces and blowout, a stellar component is not expected, while tidal models do predict a stellar component for the Leading Arm. Initial searches did not detect a stellar component of the Magellanic Stream \citep{Putman_2003, nidever_2008}. However, \citet{casetti_2014} observed 6 young stars in the Leading Arm, between $10$ and $40$ kpc from the Sun, as well as one low radial velocity O6V star at $\sim$$40$ kpc, providing evidence for recent star formation within the Leading Arm. Thus, the properties of the Leading Arm can be inferred from the stellar populations that are observed within it.

In addition to the Magellanic Stream and Leading Arm, the Magellanic outskirts are of great interest \citep[e.g.,][]{saha2010, mackey2016, nidever17, choi2018a, choi2018b, mackey2018, nidever2019, MagES1, wagner21, Petersen2022}. Studying the outskirts of the LMC can reveal the extent of Magellanic stellar populations in the sky as well as the star formation history of these regions \citep{nidever17}. Furthermore, surveys around the LMC can help identify dwarf galaxies bound to the LMC/SMC \citep{wagner16}, stellar streams neighboring the Clouds \citep{wagner21}, and faint star clusters at large physical distances \citep{nidever17}. Kinematic information, combined with chemical information, from the periphery of the Clouds can constrain the evolutionary history of the LMC-SMC-MW system \citep{munoz23}. \citet{MagES3} used data from fields between $\sim$$8^\circ$ and $\sim$$11^\circ$ from the center of the LMC to constrain the kinematics of the LMC-SMC-MW system. They also observed a mean metallicity of [Fe/H]\,$\approx-1$ across the 17 LMC fields analyzed. \citet{Petersen2022} showed that the LMC halo extends out to 30$\degr$ from the LMC center.

The \textit{James Webb Space Telescope} (JWST) provides a unique opportunity to study the stellar component of the LMC outskirts and Magellanic Leading Arm. The high resolution of the JWST, combined with its ability to search for faint stars out to $m\sim 27$ mag \citep[with S/N\,$\approx 35$ for a 3000 s exposure in F200W; e.g., Fig. 6--8 of][]{pearls_overview}, allows us to search for late-type stellar populations within the Magellanic system.

In this paper, we use JWST observations of the SMACS J0723.3-7327 galaxy cluster to search for stars that belong to the Magellanic system. Such stars likely belong to the Magellanic outskirts, given that the field of SMACS J0723.3-7327 is located between two of the 17 fields observed by \citet{MagES3} (see Figure \ref{fig:lmc_map} for a diagram of its location relative to the Magellanic system). In Section \ref{sec2} we describe the observations, the data reduction process, and the source detection methods. In Section \ref{sec3} we describe the SED-fitting methods used. In Section \ref{sec4} we describe the results from SED-fitting and the associated uncertainties. In Section \ref{sec5} we provide, in Table \ref{candidates}, a list of 71 candidate stars that are within the expected distance range of the Magellanic system and we compare these candidate stars to theoretical isochrones to estimate metallicities and ages. We conclude with a summary and prospects for future studies in Section \ref{sec6}. We express all magnitudes in the AB system \citep{abmag_ref} unless otherwise noted.

\section{Observations} \label{sec2}

\subsection{ERO and PEARLS Observations} \label{sec21}

The primary JWST data used in this study are NIRCam images of the SMACS J0723.3-7327 (hereafter SMACS\,0723) field from the JWST Early Release Observations (ERO\,2736, P.I. K. Pontoppidan). For comparison, we also use NIRCam images of three fields from the Prime Extragalactic Areas for Reionization and Lensing Science (hereafter PEARLS) survey (GTO\,1176, P.I. R. Windhorst; GTO\,2738, P.I. R. Windhorst \& H. Hammel): the North Ecliptic Pole Time Domain Field (hereafter NEP\,TDF), the IRAC Dark Field (hereafter IDF), and the MACS J0416.1-2403 (hereafter MACS\,0416) galaxy cluster. We refer the reader to \citet{Pontoppidan_2022} for more details on the SMACS\,0723 observations and \citet{pearls_overview} for details about the PEARLS program. The PEARLS images used cover the first two spokes of the NEP\,TDF, the first two epochs of the IDF, and all three epochs of MACS\,0416. The NEP\,TDF and IDF are medium-deep blank fields in the North Ecliptic Pole region and MACS\,0416 is a galaxy cluster at $z=0.4$ \citep{macs_z}. SMACS\,0723 is a galaxy cluster at $z=0.39$ \citep[e.g.,][]{ebeling_smacs_z, relics3, pascale_z, mahler_z}. SMACS\,0723 and the North Ecliptic Pole have roughly similar angular separations from the Galactic bulge of $75\degr$ and $95.8\degr$, respectively.\footnote{The galactic coordinates for SMACS\,0723 and the NEP\,TDF are $l=285.0\degr$, $b=-23.76\degr$ and $l=95.8\degr$, $b=33.6\degr$, respectively.}

The SMACS\,0723 field was observed on 2022 June 7 in the F090W, F150W, F200W, F277W, F356W, and F444W NIRCam filters, with an exposure time of 7537 s in each band. These observations reach a $5 \sigma$ AB magnitude limit of $\sim$$29.5$ mag for point sources with a $0 \farcs 08$ aperture radius according to the JWST Exposure Time Calculator (ETC). Module B of NIRCam is centered on the galaxy cluster, while module A is the noncluster module. The coordinates of SMACS\,0723 are RA=$07^h \ 23^m \ 19.5^s$, DEC=$-73\degr \ 27\arcmin \ 15\farcs 6$. 

Spoke 1 of the NEP\,TDF was observed with JWST/NIRCam on 2022 August 26 and 2022 September 14 with an observation time of 2920 s in the F090W, F115W, F410M, and F444W bands, and an observation time of 3350 s in the F150W, F200W, F277W, and F356W bands for each pointing; spoke 2 was observed on 2022 November 26-27 with the same exposure parameters. The NEP\,TDF has coordinates of RA=$17^h \ 22^m \ 47.9^s$, DEC=$+65\degr \ 49\arcmin \ 21\farcs 5$ \citep{pearls_overview}. Epoch 1 of the IDF was observed by NIRCam on 2022 July 8 with an observation time of 3157 s in the F150W, F200W, F356W, and F444W filters; epoch 2 was observed on 2023 January 6 with the same exposure parameters. The coordinates of the IDF are RA=$17^h \ 40^m \ 8.5^s$, DEC=$+68\degr \ 58\arcmin \ 27\farcs 0$ \citep{pearls_overview}. MACS\,0416 was observed by NIRCam for three epochs with, for each epoch, an observation time of 3779 s in F090W, F115W, F410M, and F444W, and an observation time of 2920 s in F150W, F200W, F277W, and F356W. Epoch 1 was taken on 7 October 2022, epoch 2 on 29 December 2022, and epoch 3 on 10 Februrary 2023. The coordinates of MACS\,0416 are RA=$4^h \ 16^m \ 8.9^s$, DEC=$-24\degr \ 4\arcmin \ 28 \farcs 7$ \citep{pearls_overview}. The NEP\,TDF, the IDF, and MACS\,0416 reach approximately 29.0 mag at $5 \sigma$ for point sources in F200W with an aperture radius of $0 \farcs 08$ \citep{pearls_overview}.

In addition to JWST NIRCam data of the SMACS\,0723 cluster, the MACS\,0416 cluster, the NEP\,TDF, and the IDF, we also use public \textit{Hubble Space Telescope} (HST) data of SMACS\,0723 from the ``Reionization Lensing Cluster Survey" (RELICS) program (GO\,14096, P.I. D. Coe). RELICS provides HST images and catalogs, of which we use the combined ACS and WFC3/IR catalog \citep{relics2, relics1}, with magnitudes in the F435W, F606W, F814W, F105W, F125W, F140W, and F160W filters. These observations have $5\sigma$ point source AB magnitude limits between 26.0 mag and 27.6 mag \citep{relics3}, and provide useful information at optical wavelengths for objects detected in the JWST data. For stars, these shorter wavelengths are critical for distinguishing between the hottest spectral types. These data are used in Section \ref{sec:hst_pickles} as consistency checks, but are not used in the final analysis. We matched the positions of stars and galaxies in the NIRCam images to those of the RELICS images. Given the one sigma astrometric residuals of $\sim$$148$ milliarcseconds between the two data sets, we found only one star to have exhibited possible proper motion of more than 3$\sigma$ and therefore concluded that proper motion is not significant between the HST and JWST observations.

\subsection{Data Reduction} \label{sec22}

All NIRCam \texttt{uncal} data were retrieved from the Mikulski Archive for Space Telescopes (MAST). For all stages of the JWST calibration pipeline, data from SMACS\,0723, the first epoch of the NEP\,TDF, the IDF, and MACS\,0416, were reduced using version 1.7.2 of the pipeline and \texttt{pmap\_0995}. Version 1.8.4 and \texttt{pmap\_1017} were used for epoch 2 of the NEP\,TDF/IDF. Finally, version 1.8.4 and \texttt{pmap\_1027} were used to reduce epoch 2 and \texttt{pmap\_1041} was used for epoch 3 of MACS\,0416. The difference between the various pipeline versions was minimal as no major NIRCam reference file updates were released between \texttt{pmap\_0995} and \texttt{1041}. The data was then processed through Stage 1 (\texttt{detector1pipeline}) of the JWST calibration pipeline, with a snowball flagging step performed before the jump detection step. Snowballs are caused by intense cosmic rays and are dealt with by expanding the data quality flags around such cosmic rays. C. Willott's \texttt{dosnowballflags.py} algorithm is used to subtract snowballs.\footnote{\url{https://github.com/chriswillott/jwst}} Next, wisp templates, which were constructed by CNAW from all publicly available NIRCam SW images, were subtracted from the NRCA3, NRCA4, NRCB3, and NRCB4 detectors for the F150W and F200W filter exposures (see \citealt{robotham_wisp} for an alternative method of removing wisps). Wisps are also present in F090W and F115W in the NEP\,TDF, but there is presently not enough public data that contains these wisps to make a high-S/N template. Following the wisp subtraction, the rate files were reduced with the default Stage 2 pipeline (\texttt{calwebbimage2}). After this, a correction for 1/f noise was applied using the ProFound package \citep{robotham_profound}, which also performed sky subtraction. Both an overall sky level and a sky gradient were fit with ProFound and subtracted from each image (see \citealt{pearls_overview} for more details). Finally, the data were astrometrically aligned and drizzled to mosaics with a pixel size of $0 \farcs 03$ using the JWST Stage 3 pipeline (\texttt{calwebbimage3}). SMACS\,0723 data was relatively aligned, whereas NEP\,TDF, IDF, and MACS\,0416 data were aligned to Gaia DR3.

\subsection{Photometry} \label{sec23}

\begin{figure*}[t]
\centerline{\fig{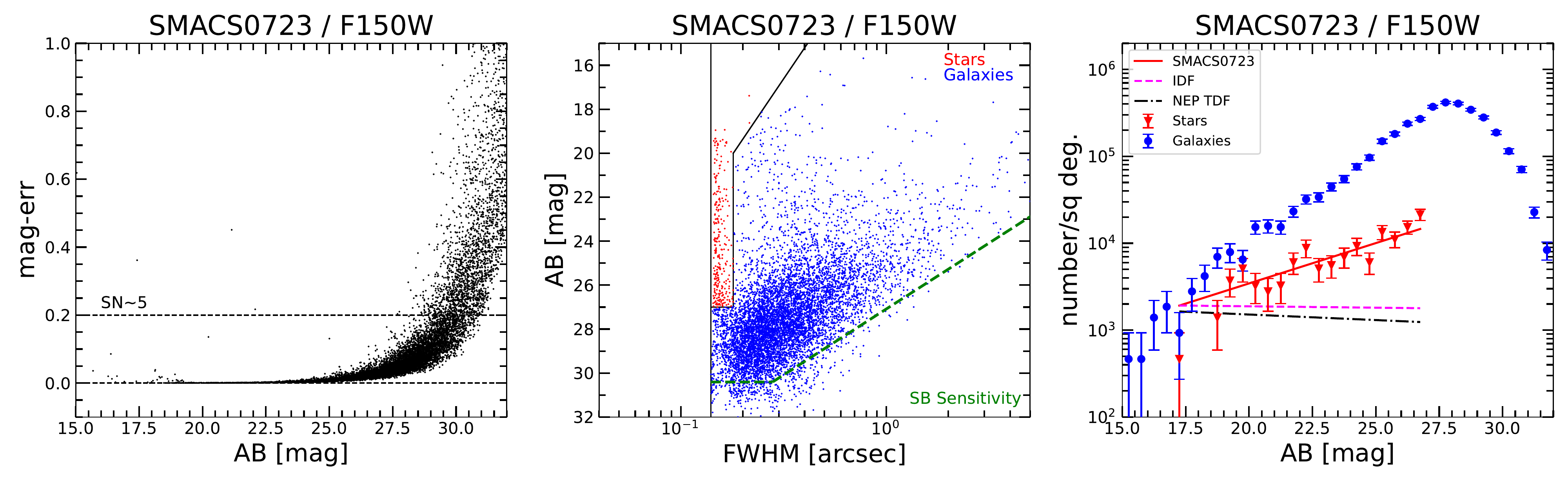}{1.10\textwidth}{}}
\vspace{-0.5cm}

\caption{(left) \texttt{SourceExtractor} \texttt{MAGERR\_AUTO} vs \texttt{MAG\_AUTO} for the F150W NIRCam filter, with a horizontal dashed line establishing the uncertainty needed for a source to have a signal-to-noise ratio of 5. (middle) \texttt{MAG\_AUTO} vs \texttt{FWHM\_IMAGE}, with stars plotted in red based on the FWHM and magnitude separation criteria. The green dashed line shows the point source and surface brightness sensitivity limit for the corresponding image (for details, see \citealt{pearls_overview}). (right) Star (red) and galaxy (blue) counts per square degree vs \texttt{MAG\_AUTO}. Best-fit lines of the star counts are plotted for SMACS\,0723, the NEP\,TDF, and the IDF for comparison. The slope of the best-fit line for SMACS\,0723 stars is greater than that of the other fields---there is an excess of faint stars.}
\label{elephant}
\end{figure*}

We use \texttt{SourceExtractor} \citep{sextractor} for source detection in the NIRCam images. First, a set of PSF-matched (to F444W) images is created. These are made by using \texttt{pypher} \citep{pypher} to calculate the necessary convolution kernel to match each filter's PSF to that of the F444W PSF, where the PSFs are simulated using \texttt{WebbPSF} \citep{webbpsf}. This convolution results in all PSFs having approximately the same FWHM and shape. \texttt{SourceExtractor} is then run on the images in dual-image mode with F444W as the detection filter and with \texttt{DETECT\_MINAREA}=9, \texttt{DETECT\_THRESH}=1.5, and a 5$\times$5 pixel gaussian convolution kernel with a FWHM of 3 pixels. F444W is among the deepest of the images, and using it results in the greatest number of stars being found, so it is used as the detection image. The \texttt{MAG\_AUTO} of each source is taken to be its observed magnitude.

To select stars from the object catalogs, we use similar methods to \citet{windhorst_2011}. We require \texttt{MAG\_AUTO} $<$ 27 and that \texttt{FWHM\_IMAGE} lies within the stellar locus of the middle plot in Figure \ref{elephant}. i.e., the FWHM fell between $0 \farcs 14$ and $0 \farcs 18$ in all filters (widening for stars brighter than 20 mag). The magnitude requirement is used to avoid classifying small, faint galaxies as stars. An object is classified as a star if it fits these criteria in at least three NIRCam filters.

\section{SED Fitting}
\label{sec3}

Near-IR spectral energy distributions are fit to the stars to estimate their distances and spectral types. To conduct the SED fitting, we use the framework of \texttt{EAZY} \citep{eazy} given its flexibility at conducting this analysis, though we emphasize that we use separate stellar SED templates rather than galaxy SED templates. To do this, the maximum redshift is set to $z=0$ and the minimum filters used in the fitting is set to 4. Before fitting SEDs, the catalogs (other than the IDF, which is only used for star counts) are corrected for galactic extinction ($A_V=0.587$ mag for SMACS\,0723, $A_V=0.082$ mag for the NEP\,TDF, and $A_V=0.101$ mag for MACS\,0416).  We apply the full value of $A_V$ to better fit distant stars which have more galactic dust in front of them. The $A_V$ values are likely quite different between fields due to LMC dust in the field of SMACS\,0723, although the correction only causes an adjustment of $\sim$$0.05$ mag in F200W for SMACS\,0723. The SMACS\,0723 RELICS catalog has the same $A_V$ correction by default. We use fluxes for the \texttt{EAZY} SED fitting.

\subsection{Template Construction}

Instead of the default galaxy SED templates, we use stellar SED templates from the SpeX IRTF Spectral Library from \citet{spex2009} and \citet{spex2005}. The SpeX library covers wavelengths of $0.8 - 5.0 \ \micron$, with templates for stars with MW abundances, cooler in temperature than type A.\footnote{Tempaltes for F2V, F3V, F4V, F5V, F7V, F8V, F9.5V, G0V, G1V, G2V, G3Va, G5V, G6.5V, G8V, K0V, K1V, K2V, K3V, K5V, K7V, M0V, M0.5V, M1V, M1.5V, M2.5V, M3V, M3.5V, M4V, M5V, M6V, M6.5V, M7V, M8V, M9V, M9.5V, L1, L3, L3.5, and L5 stars are used from \url{http://irtfweb.ifa.hawaii.edu/~spex/IRTF_Spectral_Library/}.} We only fit main sequence stars because they account for most stars in the LMC field. If we assume there are $25,000$ AGB stars per 16 deg$^2$, as in \citet{wiz2011}, then there could be around 3 AGB stars in the NIRCam SMACS\,0723 data. This is insignificant compared to the number of total stars observed, and since we are focusing on the overall trend in star distances, these stars can safely be ignored. 

We then use \texttt{pysynphot} \citep{pysynphot} to compute the apparent magnitude in every NIRCam filter for each stellar SED template. The Gaia parallax of each SpeX star is then used to determine its absolute magnitude in each NIRCam filter \citep{gaia1, gaia2, gaia3}. Since spectral type for main sequence stars roughly correlates with their absolute magnitudes, the absolute magnitude of each SpeX star can be estimated given its parallax-based distance and its apparent magnitude.

\subsection{Bayesian Probabilities} \label{sec:bayes}

When fitting stellar SEDs with \texttt{EAZY}, the code is ran once for each individual template, the template is fit to every star and the spectral type, $\chi^{2}$, and observed/template SEDs are recorded. All stars with best-fit $\chi^2_{\text{red}} > 2$ are removed from the final catalog. This cut removes 21 objects from the SMACS\,0723 catalog. Most high-$\chi^2$ objects are likely either not main-sequence stars or have questionable photometry. 

To provide an estimate of the distance moduli, $\mu$, of the stars in SMACS\,0723, we measure the posterior probability distributions of the distance modulus for each star. To do this, we use the output from \texttt{EAZY}, which provides the fitting parameters for every star for each stellar SED template. Bayes' Theorem is then used:
\begin{equation} \label{bayes}
    P(\text{Template} | \text{Fluxes}) = \frac{P(\text{Fluxes} | \text{Template}) P(\text{Template})}{\sum_i P(\text{Fluxes} | \text{Template}_i)},
\end{equation}
where each term is a probability. The measured fluxes in each NIRCam filter for a given star are the ``Fluxes" and the spectra for a given spectral type (SpeX star) is the ``Template." Equation \ref{bayes} solves for $P(\text{Template} | \text{Fluxes})$ (the posterior probability), which is the probability that a star fits a particular SpeX SED template, assuming that the fluxes are accurate. $P(\text{Fluxes} | \text{Template})$ is the probability that the fluxes are correct, assuming that the star fits a particular SpeX SED template (which corresponds to a spectral type), and $P(\text{Template})$ is the probability of a random star fitting the SpeX SED template (the prior). We assume a constant prior probability model (changing this to reflect the MW IMF has almost no effect on the results) on absolute magnitude, turning Equation \ref{bayes} into the following:
\begin{equation}
    P(\text{Template} | \text{Fluxes}) = \frac{P(\text{Fluxes} | \text{Template})}{\sum_i P(\text{Fluxes} | \text{Template}_i)}.
\end{equation}

We calculate $P(\text{Fluxes} | \text{Template})$ as the product of gaussians (one per filter) centered at each predicted flux, with a standard deviation equal to \texttt{EAZY} \texttt{err\_full} (includes photometric uncertainty and template uncertainty), evaluated at the measured flux. The posterior probability for each template fitting a given star is equal to the probability that the template fits the star divided by the sum of the probabilities over all of the templates. This posterior probability distribution (as a function of spectral type, which each have a corresponding absolute magnitude) is then calculated for all stars in all filters and is converted from spectral types to absolute magnitudes. The probability distributions are then converted to distance moduli by subtracting the distribution from each filter's apparent magnitude. Therefore each distribution gives the probability that a given star is fit with a given distance modulus. These probability distributions are then linearly interpolated and all 6-8 of them (6-8 NIRCam filters) are summed together for each star. These probability density functions are normalized to an area of one, and are used to calculate the lower bound, median, and upper bound of the distance modulus for each star. This is done by measuring the 16th, 50th, and 84th percentile in the cumulative probability distribution, respectively. Thus the reported uncertainties correspond to $1\sigma$ uncertainty. Measuring the probability densities of each star allows for more complete assessment of uncertainties, reported in Table \ref{candidates}.

\subsection{HST Checks} \label{sec:hst_pickles}

To confirm the SED fits from the SpeX library of stellar SEDs, we use stellar SED templates from the optical library from \citet{pickles}. For stars in the field of SMACS\,0723 that have HST data available (there are 90 of them), the optical HST data (F435W, F606W, F814W, F105W, F125W, F140W, and F160W) and short wavelength NIRCam data (F090W, F150W, and F200W) are fit with \texttt{EAZY} by the \citet{pickles} templates (the templates cover $0.12 - 2.50 \ \micron$ in wavelength) for a variety of main sequence stars.\footnote{Available templates include those of O5V, O9V, B0V, B1V, B3V, B6V, B8V, A0V, A2V, A3V, A5V, F0V, F2V, F5V, F8V, G0V, G2V, G5V, G8V, K0V, K2V, K5V, K7V, M0V, M4V, and M5V stars from \url{https://archive.stsci.edu/hlsps/reference-atlases/cdbs/grid/pickles/}.} The Pearson correlation coefficient ($r$) was calculated between the shared templates for the two spectral libraries over their shared wavelengths, and all templates except for the M5V ($r \approx 0.76$) template agree to $r \gtrsim 0.97$. However, the M5V template is the last template available from the \citet{pickles} library, so this is to be expected. Since the \citet{pickles} templates are normalized and combine multiple stars for the creation of each template, we assume each template has the same absolute magnitude as the SpeX star with the same spectral type. The absolute magnitudes fit by the two libraries have the greatest agreement at $M_{\text{F090W}} \lesssim 7$ mag, where the standard deviation of their difference is $\sim$$0.35$ mag. This limited range of agreement is to be expected because the SpeX library offers a greater variety of late-type stellar SED templates compared to the \citet{pickles} library. Since the SpeX library leads to similar results as the \citet{pickles} library, covers more near-IR wavelengths, and has a wider variety of M-star templates, it is reasonable to use the SpeX library with all of the SMACS\,0723, NEP\,TDF, and MACS\,0416 photometry.

\section{Magellanic Star Candidate Identification} \label{sec4}

\subsection{The Distance Modulus}

\begin{figure*}[t]
\centerline{
\fig{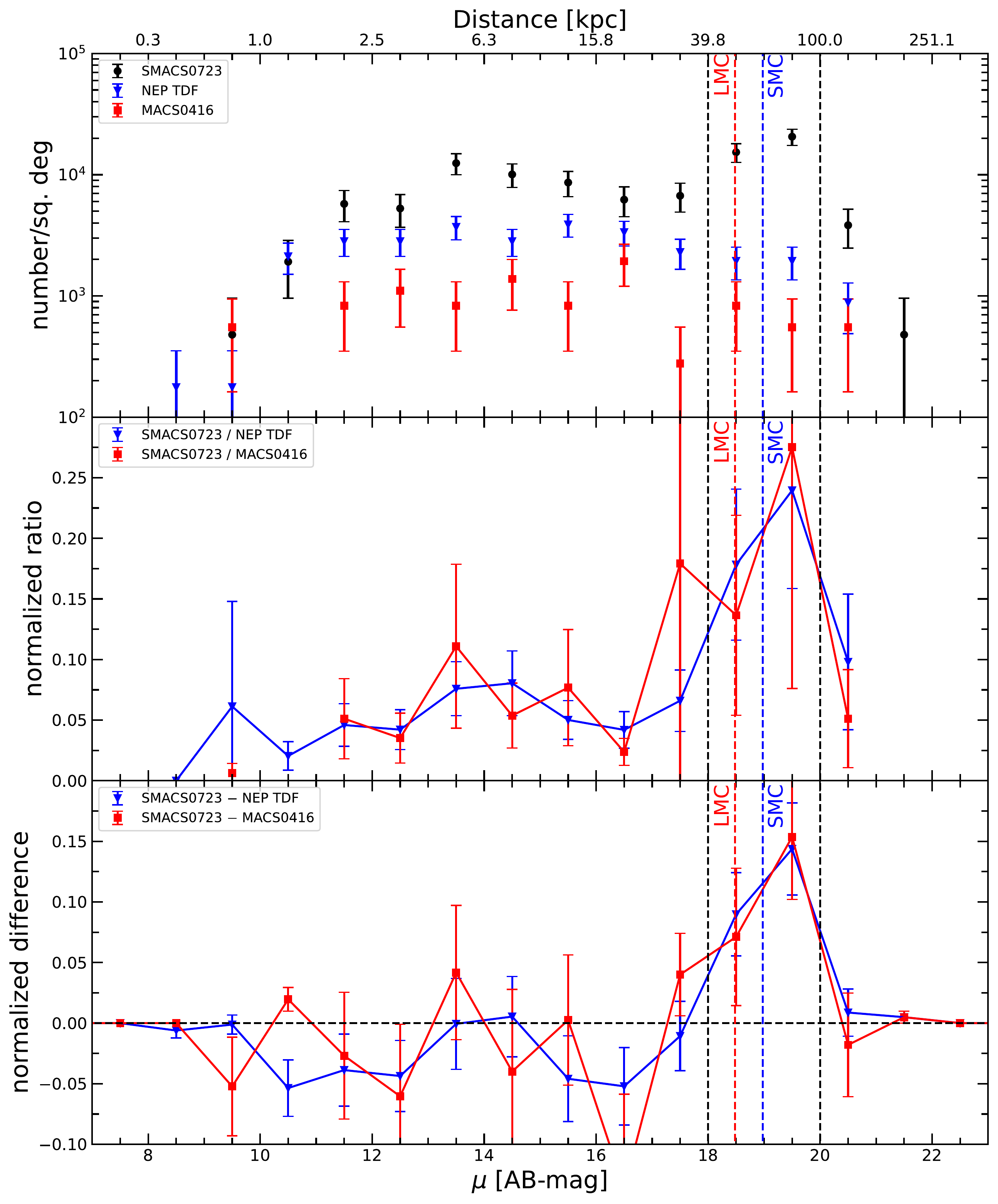}{1.0\textwidth}{}}
\vspace{-0.5cm}
\caption{Distribution of distance moduli, $\mu$, (calculated from the 50th percentile in the distance modulus PDF) for stars in the SMACS\,0723, the NEP\,TDF, and the MACS\,0416 fields. (top) Binned distance moduli for stars in each field. (middle) Ratio of stars found at each distance in SMACS\,0723 compared to the NEP\,TDF and MACS\,0416, normalized to have a sum of one. (bottom) Difference in probability density between SMACS\,0723 and the NEP\,TDF/MACS\,0416 fields, where the data from each field is normalized to have a sum of one (resulting in a difference of zero between fields). The vertical black dashed lines at a distance modulus of 18 and 20 mag show the ranges of candidates taken, the vertical red dashed lines shows the distance to the LMC, and the vertical blue dashed lines show the distance to the SMC.}
\label{dmod}
\end{figure*}

We consider two ways of estimating the distribution of distance moduli of stars in the SMACS\,0723 field. First, we consider the distribution of distance estimates of each star (using the 50th percentile of the distance modulus PDF). Alternatively, all of the probability distributions of each star can be added up to form one probability distribution for each field as a whole. Figure \ref{dmod} shows the distribution of 50th percentile distance moduli for each field. The ratio of the distributions, shown in the middle plot, shows a clear surplus of stars in the field of SMACS\,0723 at $18 \lesssim \mu \lesssim 20$ mag, which corresponds to a distance between $\sim$$39.8 \ \text{kpc}$ and $\sim$$100.0 \ \text{kpc}$. We refer to the stars in this range of distances as stellar candidates, however the properties of each individual star are more uncertain compared to the population as a whole. Additionally, some candidate stars could be part of the MW Halo since the JWST photometry does not have the spectral resolution to detect precise metallicity differences between individual stars. The integrated surface density between these distance moduli is $\sim$$50,500$ stars deg$^{-2}$ in the SMACS\,0723 field, compared to $\sim$$8200$ stars deg$^{-2}$ in the NEP\,TDF. This means that the field of SMACS\,0723 has $\sim$$6.1$ times as many stars per sq. degree as the NEP\,TDF between 39.8 and 100 kpc, compared to an overall ratio of $\sim$$2.3$ and a ratio at all distances other than $18\leq\mu\leq20$ mag of $\sim$$1.9$. Overall, the probability distribution for the SMACS\,0723 field has $\sim$$7600$ more stars deg$^{-2}$ mag$^{-1}$ than the NEP\,TDF. However, between 39.8 and 100 kpc the difference is $\sim$$21,100$ stars deg$^{-2}$ mag$^{-1}$. Figure \ref{dmod} shows the difference in 50th percentile star counts between the SMACS\,0723 field and the two other fields in the bottom plot. The field of SMACS\,0723 has a clear excess between 39.8 and 100 kpc, whereas the NEP\,TDF and MACS\,0416 field have a minor excess relative to the SMACS\,0723 field at closer distances. Hence, there is a significant surplus of stars ($15.3 \sigma$ compared to the NEP\,TDF, $7.2\sigma$ compared to the field of MACS\,0416) in the field of SMACS\,0723 at this distance. The range of distances found in this paper is different compared to \citet{casetti_2014}, which observed early-type stars at a distance of up to $\sim$$40 \ \text{kpc}$. The stars observed in this paper are farther than $40$ kpc, likely because the field of SMACS\,0723 is closer in the sky to the LMC than those in \citet{casetti_2014}. The star candidates are expected to be closer to $\sim$$50 \ \text{kpc}$ because they are only $\sim$$10^\circ$ from the LMC and the field of SMACS\,0723 is near the line-of-nodes for the LMC's inclination \citep{MagES3}, unlike the stars in \citet{casetti_2014}, which were further stars in the Leading Arm.

\subsection{Contamination}

\begin{figure}[t]
    \centerline{\fig{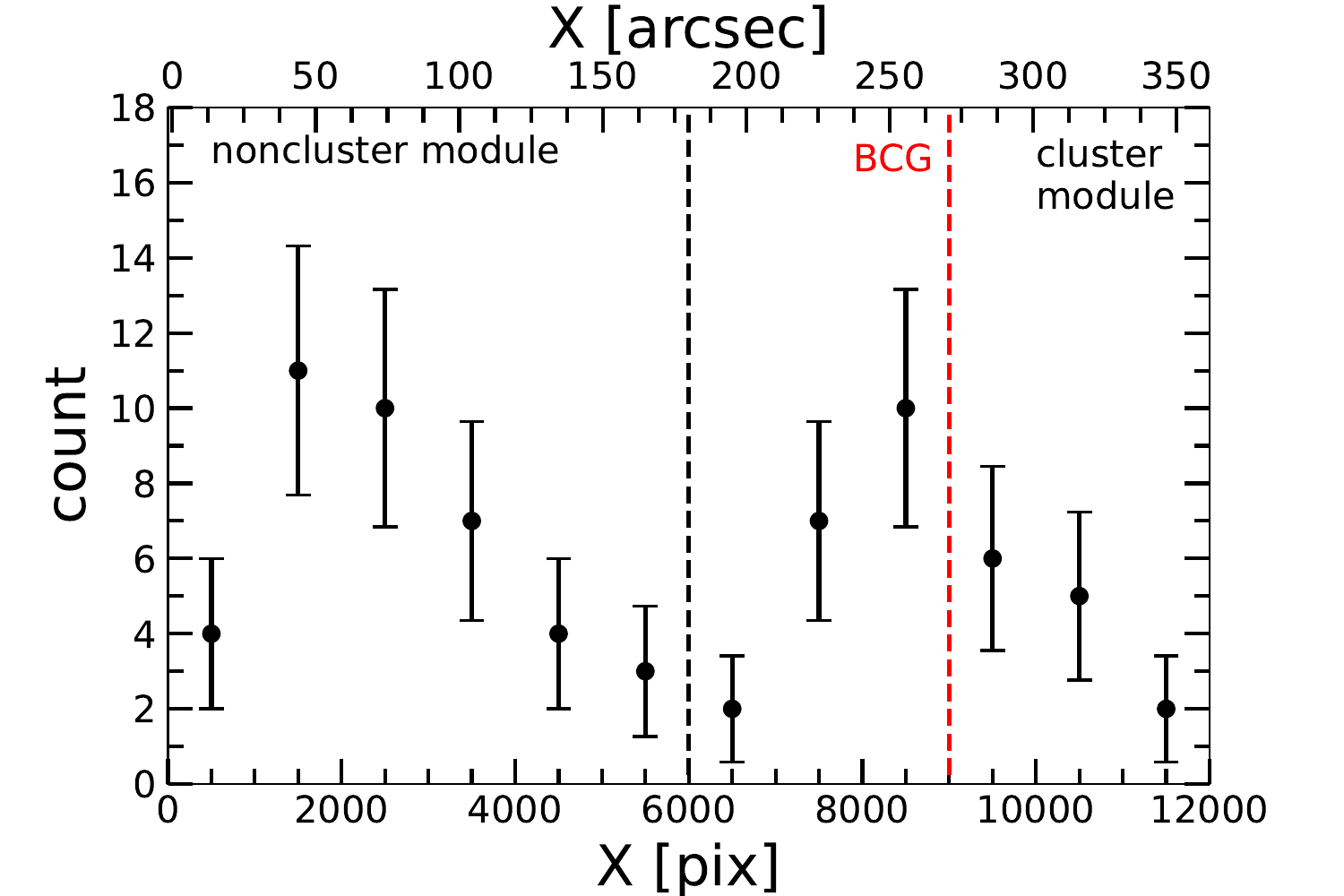}{0.5\textwidth}{}}
    \vspace{-0.5cm}
    \caption{Histogram of X-positions of candidate stars in the SMACS\,0723 mosaics. Error bars show $\sqrt{N}$ uncertainties. The noncluster module is on the left of the vertical black dashed line and the cluster module is on the right, with the position of the brightest cluster galaxy shown as a vertical red dashed line. No significant excess of candidate LMC stars around the BCG is detected.}
    \label{fig:cand_dist}
\end{figure}

Since SMACS\,0723 is a galaxy cluster, an obvious source of contamination would be dwarf galaxies or globular clusters within the galaxy cluster. We would expect foreground LMC stars to be distributed randomly throughout the SMACS\,0723 images, and not concentrated in the NIRCam module that contains the SMACS\,0723 cluster center.  As a test we counted the numbers of stellar candidates in the NIRCam module centered on the cluster (32 stellar candidates) and offset from the SMACS 0723 cluster (39 stellar candidates), obtaining numbers which are less concentrated towards the cluster center. Figure \ref{fig:cand_dist} shows the number of candidates as a function of X-coordinate. Since the mosaics are not rotated to be north-up, the X-coordinate is equivalent to the lengthwise distance along the NIRCam field of view. As expected, there is a dip in star counts close to the module gap. There is an excess of candidate stars near the BCG, but the noncluster module has a similar spike in stars elsewhere, so we conclude that contamination in this catalog is unlikely. The centroid of LMC candidates for each NIRCam module is $\lesssim$$250$ pixels from the center of each module. Thus the candidates are not skewed a certain direction and are randomly distributed as expected.

In addition to the spatial distribution of candidate stars in the field of SMACS\,0723, the inclusion of MACS\,0416 (which has approximately the same redshift at $z=0.4$, as SMACS\,0723 at $z=0.38$) data further corroborates that there is little to no contamination. In particular, Figure \ref{dmod} shows that there is no excess of stars in the field of MACS\,0416 at $18 \leq \mu \leq 20$ mag. Thus the excess in point sources at $18 \leq \mu \leq 20$ mag is not caused by contamination from the SMACS\,0723 galaxy cluster.

\subsection{Uncertainty in $\mu$}
There are various contributions in the uncertainty budget of the calculated distance moduli, including photometric uncertainties, SpeX calibration uncertainties, and SED fit uncertainties, which we account for. Unaccounted for uncertainties include distance-dependent dust extinction as well as metallicity differences between the stars observed and the \texttt{EAZY} templates used.  Photometric uncertainties in the NIRCam data and the SpeX calibration only contribute $\lesssim$$3\%$ uncertainties to the distance moduli. This was verified using Monte Carlo methods: the observed magnitudes were resampled based on their \texttt{SourceExtractor} uncertainties 1000 times for each star, the 1000 probability distributions for each star were averaged together, and the 50th percentile in the PDF of distance moduli was recalculated to be $\lesssim$$3\%$ of the distance modulus calculated with only the observed magnitudes. The calibration for the IRTF spectral library is within a few percent \citep{spex2009}. For a conservative estimate on uncertainty in the SED templates, the 2MASS photometry typically has uncertainties of $\lesssim$$0.25$ mag, so we add this number in quadrature to the final upper and lower uncertainties. The uncertainty in parallax for each of the SpeX stars used is $\lesssim$$0.01$ mag, and is therefore negligible. Finally, the SED fit uncertainties are described in section \ref{sec:bayes} and generally have the greatest contribution to the overall uncertainty in $\mu$. Thus the 2MASS photometric uncertainty and the SED fit uncertainties are used to calculate the final uncertainties in $\mu$, reported in Table \ref{candidates}. 

\subsection{Gaia Stars}

A Gaia DR3 \citep{gaia1, gaia2, gaia3} query\footnote{The Gaia DR3 catalog can be queried at \url{https://gea.esac.esa.int/archive/}.} was performed in each of the fields studied in this paper to check if any of the detected stars had measured parallaxes. A handful of the JWST stars from each field have parallaxes and proper motions, but the uncertainties are large enough to render them unhelpful. As a separate check, a Gaia DR3 query with a larger area was undertaken around the field of SMACS\,0723 and the NEP\,TDF to check for an excess of stars at the distance of the LMC. However, due to the shallower depth of Gaia, no significant excess of low-parallax stars nor stars with small proper motion characteristic of the LMC was found. Thus Gaia data alone is not enough to detect an excess of stars around the field of SMACS\,0723 at the distance of the LMC.

\section{Magellanic Star Candidate Properties} \label{sec5}

\begin{figure*}[t]
\centerline{\fig{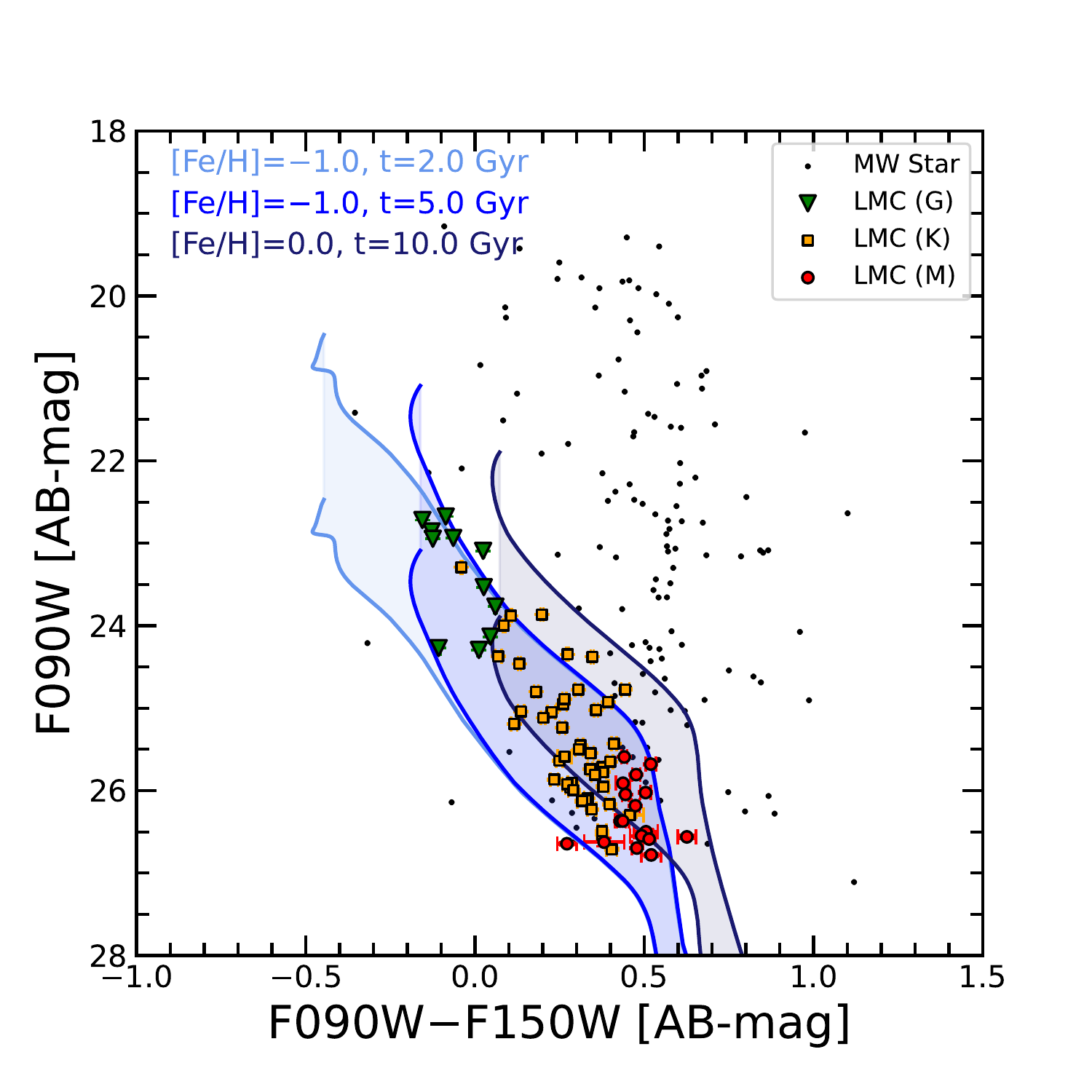}{0.55\textwidth}{}
          \fig{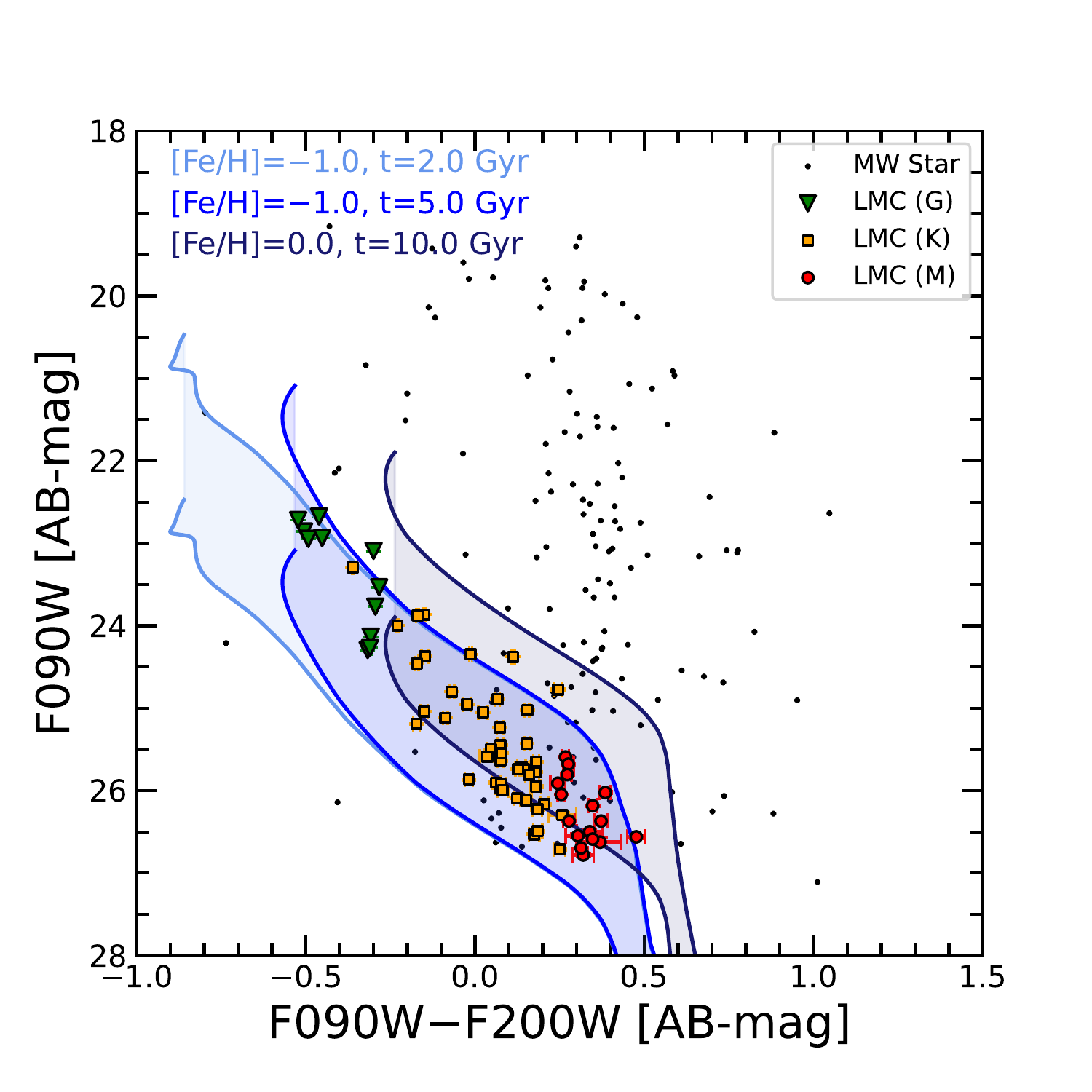}{0.55\textwidth}{}}
\vspace{-1.5cm}
\centerline{\fig{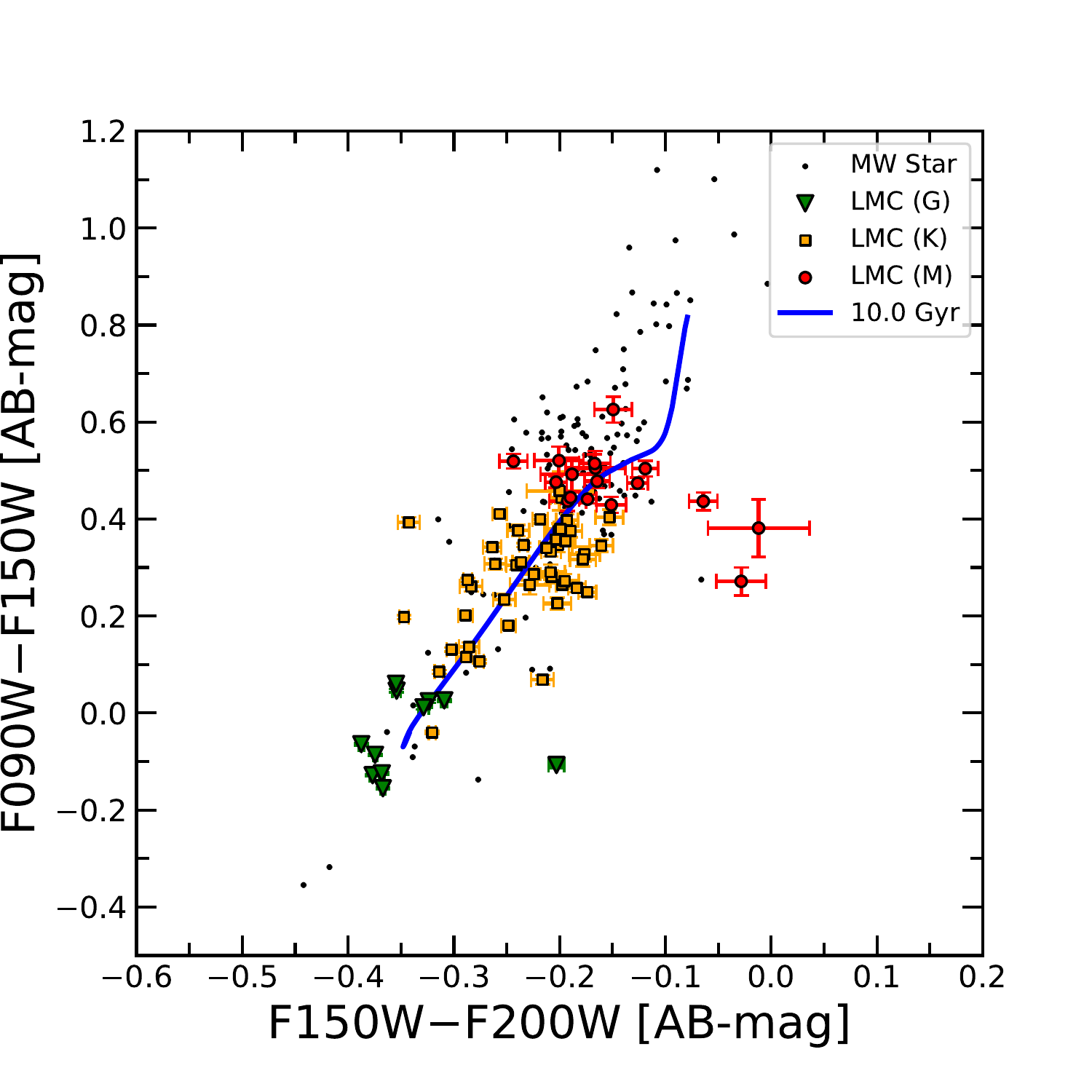}{0.7\textwidth}{}}
\vspace{-0.5cm}
\caption{(top) Color-magnitude diagrams including the 71 candidate stars (with colored shapes corresponding to different spectral types) and the rest of the detected stars (black dots). Plotted on them are shaded isochrones of various ages and metallicities for distance moduli between 18 and 20 mag. (bottom) Color-color diagram with the 71 candidate stars as well as an isochrone of $t=10.0$ Gyr and [Fe/H]\,$=-1.0$. A few candidates typed as G stars must be younger than this age.}
\label{CMD}
\end{figure*}

\subsection{Age and Metallicity Estimates}

We now set out to estimate the approximate ages and metallicities of the LMC candidate stars. When selecting star candidates that are part of the Magellanic system, we required that $18 \leq \mu \leq 20$ mag and $\chi^2_{\text{red}} < 1$. This range of distances allows for the detection of nearby stars in the Leading Arm (possibly around $40$ kpc) as well as farther stars in the Magellanic outskirts closer to $50$ kpc. The smaller $\chi^2_{\text{red}}$ value selects only the candidates that fit their stellar template the best. These LMC stellar candidates are presented in Table \ref{candidates}.

Figure \ref{CMD} shows all of the stars on color-magnitude diagrams, with isochrones corresponding to stellar populations between $\mu=18$ mag and $\mu=20$ mag. The isochrones are created using MESA Isochrones \& Stellar Tracks data \citep{MIST1, MIST2, MIST3, MIST4, MIST5}. The isochrones assume [Fe/H]\,$= -1.0$ and $v/v_{crit}=0.4$ rotation rates. We also generate isochrones of [Fe/H]\,$=0.0$ to represent MW stars. Most of the candidate stars (colored shapes) fall in the range of the isochrones, reflecting that the candidates form their own stellar population. The other detected stars (black dots) that fall around the isochrones generally have a distance modulus close to that of the LMC, but outside the range of distance moduli used to select candidates.

The characteristics of the isochrones reflect the metallicity of this stellar component in the LMC. The implied metallicities seem to be consistent with [Fe/H]\,$=-1.0$. This metallicity is consistent with that of stars observed in the LMC outskirts \citep{MagES3}. Many candidates have colors and magnitudes that are not consistent with having [Fe/H]\,$=0.0$, so are not likely to be MW stars. Some other candidates presented could be MW stars, but that information could only be distinguished via precise metallicity or proper motion measurements. The stars that are almost considered candidates, with $\mu \lesssim 18$ mag or $\mu \gtrsim 20$ mag, could also be LMC stars, but we cannot be certain of this since the difference between the SMACS\,0723 field and the NEP\,TDF at that physical distance is not significant. 

Figure \ref{CMD} also shows a color-color diagram with the candidate stars as colored shapes and the other detected stars as black dots. In general, the candidates are closer to the lower-left (hotter) than the bulk of observed stars, Many non-candidate stars are closer to the upper right (type M). Additionally, the isochrone of $10$ Gyr (with [Fe/H]\,$=-1.0$) shows that at least a few candidates are inconsistent with the ages of MW halo stars.

Of all 204 stars identified in the field of SMACS\,0723, only two are fit as type F stars. These two stars have relatively blue colors, and can be seen as black dots in the the bottom left corner of the color-color diagram in Figure \ref{CMD}. These stars are the bluest two stars in the color-magnitude diagrams in Figure \ref{CMD}. Both stars have HST data available. The first of the two type F stars was fit as F9.5V with $\mu = 20.2^{+0.8}_{-0.3}$ mag using the SpeX templates and G6.5V with $\mu = 19.9^{+0.3}_{-0.3}$ mag using the \citet{pickles} templates. The second star was fit as type F5V with both libraries, and had $\mu = 18.0^{+0.4}_{-0.7}$ mag using the SpeX templates and $\mu = 18.5^{+0.3}_{-0.6}$ mag using the \citet{pickles} templates. Thus both stars are slightly outside of the criteria used for selecting Magellanic system candidates. However, it is still possible that these stars are part of the LMC outskirts. If this is the case, then there could be recent star formation within the LMC outskirts from $<5.0$ Gyr ago.

\subsection{Absolute Magnitudes and Masses}

\begin{figure}[t]
    \centerline{\fig{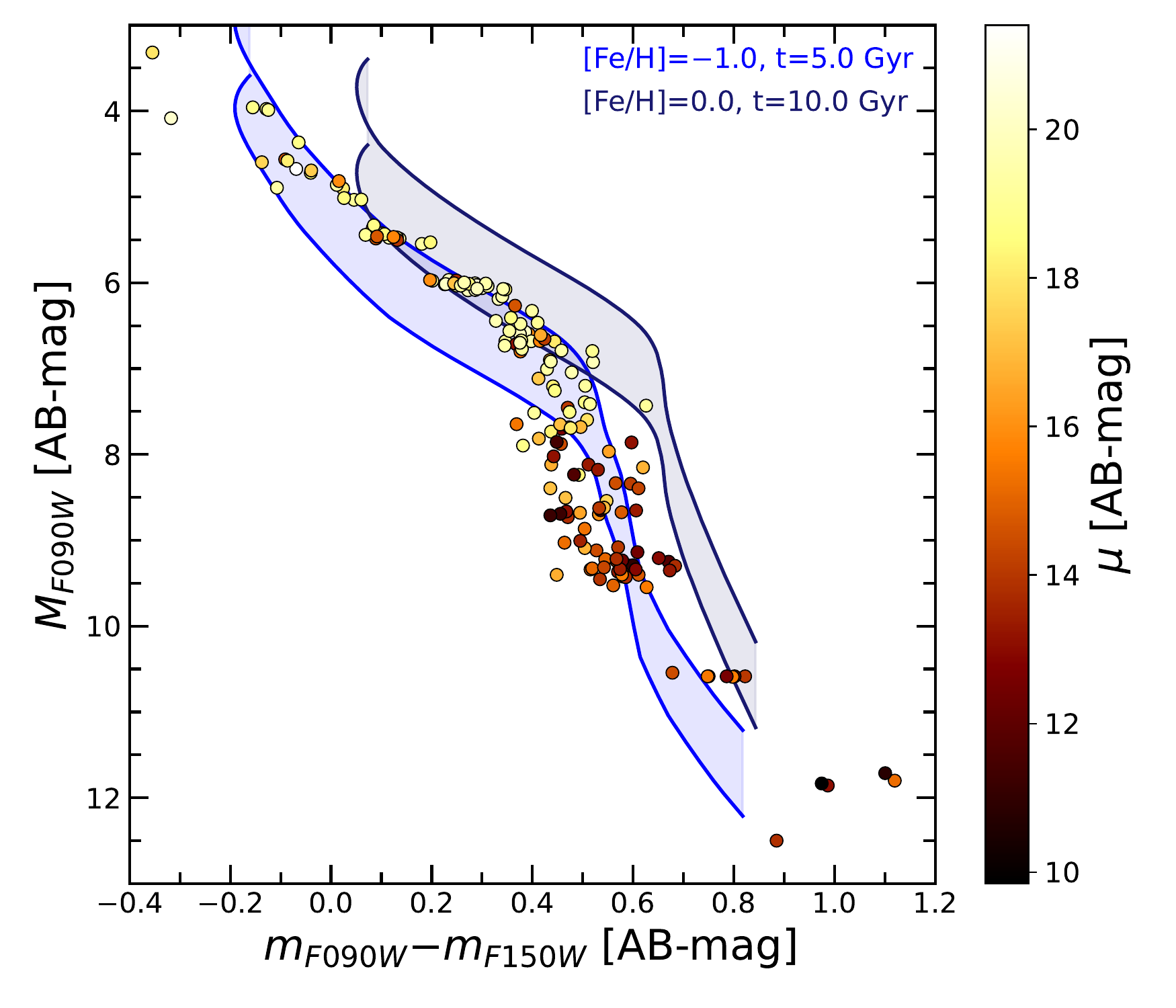}{0.5\textwidth}{}}
    \vspace{-0.5cm}
    \caption{HR Diagram of all stars observed in the field of SMACS\,0723. Distance modulus ($\mu$) is indicated by the color-scheme, and isochrones are plotted to indicate the expected positions of an old MW stellar population and a younger LMC population.}
    \label{fig:HR}
\end{figure}

\begin{figure}[t]
    \centerline{\fig{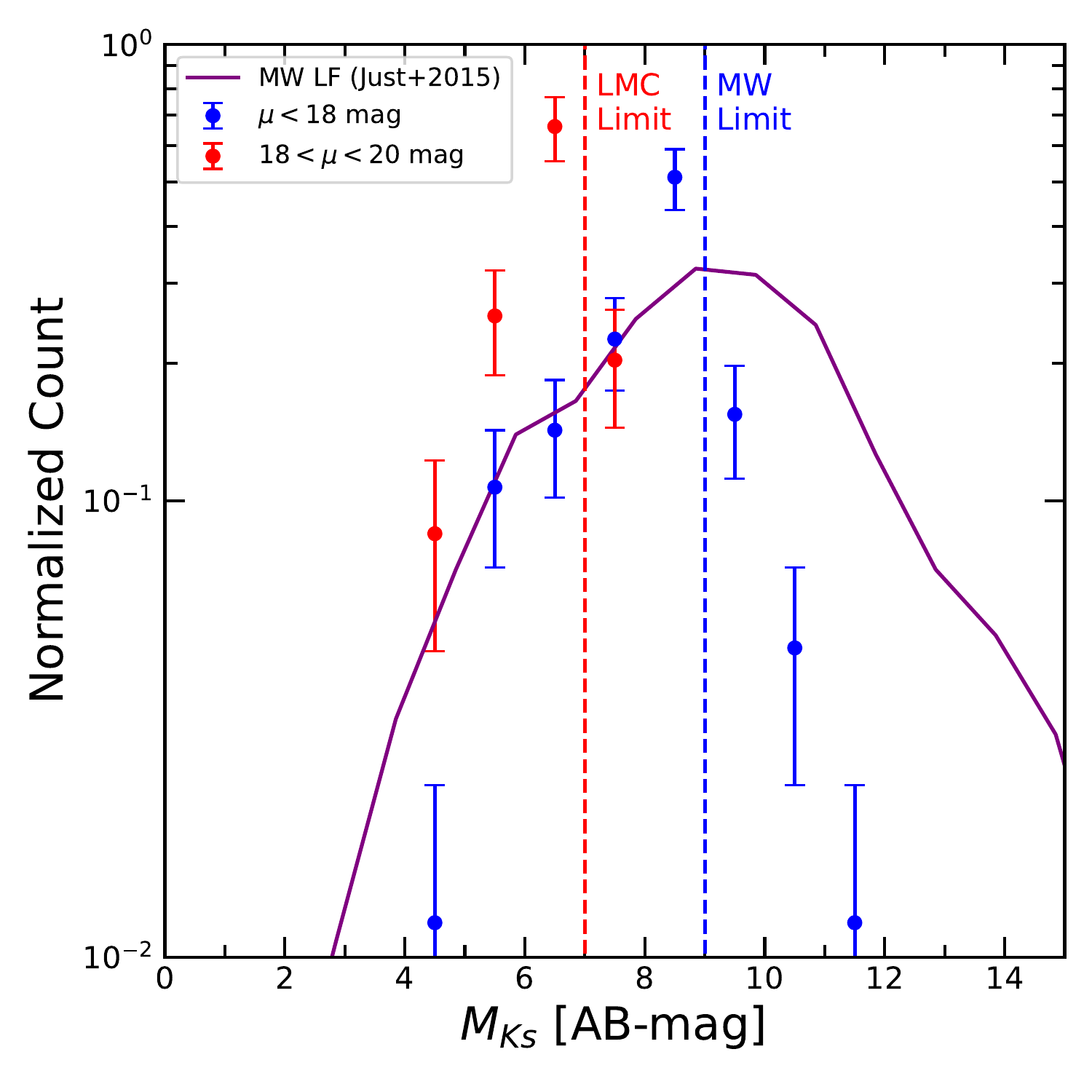}{0.5\textwidth}{}}
    \vspace{-0.5cm}
    \caption{Histogram of absolute magnitudes (in the 2MASS K$_{s}$ bandpass) for both the LMC candidate stars (in red) and the remaining MW stars (in blue). The MW luminosity function measured in \citet{just2015} is plotted in purple. Dashed lines indicate the incompleteness limit of our sample (due to the magnitude cutoff of $m<27$ mag and the distance cutoffs for each population).}
    \label{fig:LF}
\end{figure}

We now use the measured absolute magnitudes of the stars in the field of SMACS\,0723 to evaluate the properties of each stellar population (MW and LMC) as well as estimate stellar masses. Figure \ref{fig:HR} shows the HR diagram for the stars observed in this study. Bluer stars tend to be fit by either [Fe/H]\,$=-1.0$ or [Fe/H]\,$=0$. However, redder stars are better fit by the [Fe/H]\,$=-1.0$ model.

We now use the measured absolute magnitudes to compare our observed stellar luminosity function (LF) to that in Figure 6 of \citet{just2015}. Figure \ref{fig:LF} shows the luminosity function for the set of MW stars, the set of LMC stars, and the \citet{just2015} results. The data is normalized to have a sum of 1 for stars brighter than the completeness limits. \citet{just2015} plots nearly all stars in the $20$ pc sphere around the Sun, whereas the field of SMACS\,0723 is a single pointing outside the MW disk. The LF observed for the LMC candidates is $\sim$$2$ mag brighter than that of the closer MW stars ($\mu < 18$ mag). Additionally, the LF observed for the MW stars matches up better with the \citet{just2015} LF, although it is incomplete for $M_{K_s} \gtrsim 9$ mag.

\begin{figure}[t]
    \centerline{\fig{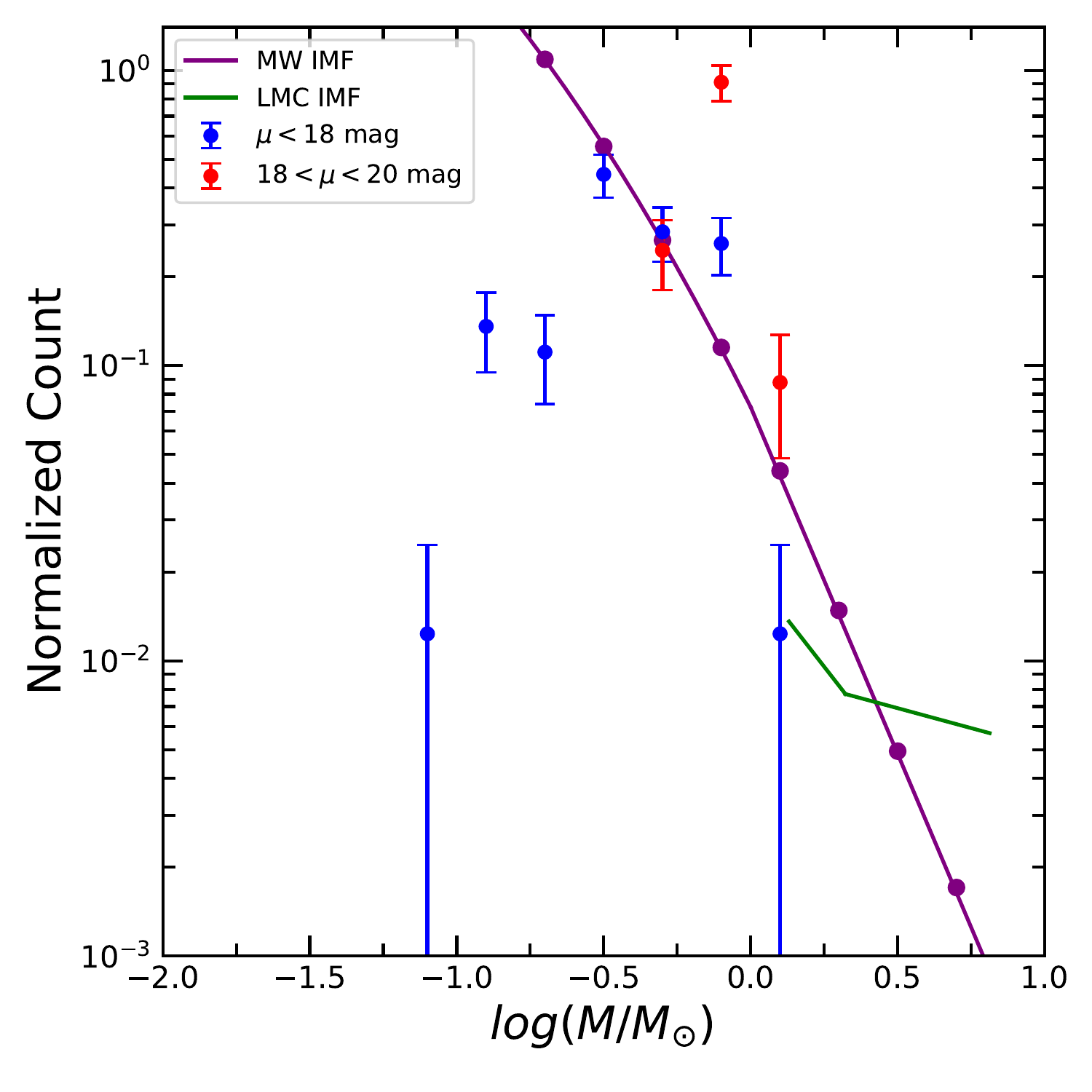}{0.5\textwidth}{}}
    \vspace{-0.5cm}
    \caption{Histogram of masses observed in each stellar population (MW, blue: $\mu < 18$ mag; LMC, red: $18 < \mu < 20$ mag). The MW initial mass function (IMF) is plotted in purple using the \citet{salpeter1955} IMF for $\log(M/M_{\odot}) \geq 0$ and the \citet{chabrier2003} IMF for $\log(M/M_{\odot}) \leq 0$. The IMF observed in \citet{sirianni2000} for LMC star clusters is plotted in green for $0.13 \leq \log (M/M_{\odot}) \leq 0.81$. Graphs are normalized to a sum of 1 for masses greater than the completeness limits. Any mass bins with counts less than $10^{-2}$ have zero stars in them and are not plotted.}
    \label{fig:IMF}
\end{figure}

Using the absolute magnitudes derived for each star, we then estimate stellar masses. To do this, we measure the K$_{s}$ absolute magnitude for each SpeX template using \texttt{pysynphot} \citep{pysynphot}. This allows us to use the mass relations in \citet{henry1993}. The results of this analysis are plotted in Figure \ref{fig:IMF}. We compare the observed IMFs to the \citet{salpeter1955} and \citet{chabrier2003} MW IMF, plotted in purple, as well as the \citet{sirianni2000} LMC IMF, plotted in green. The results are not complete for lower mass stars, but the MW star masses ($\mu < 18$ mag) show agreement for $-0.6 \leq \log(M/M_{\odot}) \leq 0.2$. The LMC stellar masses ($18 < \mu < 20$ mag), on the other hand, are more biased towards higher mass stars and do not fit the MW IMF well. The LMC IMF is not complete for low masses and so does not overlap with our data.

\subsection{Stellar Density}

Taking the areal coverage of SMACS\,0723 to be $\sim$$0.002$ sq. deg., our 71 candidates would lead to a stellar density of $\sim$$34,000 \ \text{deg}^{-2}$. This corresponds to a spatial density at a distance of 50 kpc (between 40 and 100 kpc from the Sun) of 740 stars kpc$^{-3}$. This spatial density will further help models constrain the orbital history of the Magellanic system around the Milky Way. In particular, measuring the stellar density profile of the LMC can help reveal faint features of the LMC other than the primary elliptical profile \citep{massana20}. With shallower data, the surface brightness profile of the LMC was not detected beyond galactocentric radii of $\sim$$8\degr$ \citep{massana20}; deeper data at farther radii, such as with the field of SMACS\,0723, can reveal a more complete LMC profile. Compared to \citet{MagES3}, we observe many more faint stars, since the angular density of Gaia stars around this area of the LMC outskirts (around the field of SMACS\,0723 and two \citealt{MagES3} fields) is only $\sim$$45$ stars/deg$^{-2}$. The ages observed are inconsistent with $t=10$ Gyr because there are a few type G stars (plotted as stars and triangles in Figure \ref{CMD}) that are likely relatively young. This suggests that the group of stars observed has a similar dynamical age as the body of the LMC.

The expected number of main sequence turnoff (MSTO) stars at an angular distance of $10\degr$ from the LMC center is around $600$ stars/deg$^2$ \citep{nidever2019}. Since our data goes deeper, it is difficult to directly compare our results with other studies. However, if we take the number of G stars (since the MSTO in \citealt{nidever2019} go to $\sim$$9$ Gyr) observed in the field of SMACS\,0723 at near LMC distance, we find $\sim 5300$ stars/deg$^2$. If instead we consider stars with $2.5$ mag $\leq M_{\text{F090W}} \leq$ $4.5$ mag, we find a stellar density of $\sim 1900$ stars/deg$^2$. These stellar densities are somewhat different due to using different methods to select stars. This could mean that there are more LMC stars in the direction of SMACS\,0723, or that the comparison is inexact.

\section{Conclusion} \label{sec6}

We observe a total of 71 candidate stars that are between 39.8 and 100 kpc from the Sun, according to their SED-derived distance moduli. These stars are in a field that is $\sim$$10^\circ$ or $\sim$$9$ kpc from the LMC. We present information about these candidates in Table \ref{candidates}. The candidate stars could be in the outskirts of the LMC or part of the MW halo. They could also be part of the Leading Arm, but the distances are too uncertain to separate the Leading Arm from the LMC outskirts. The stars lie around isochrones of this distance with [Fe/H]\,$=-1.0$, which is the expected metallicity for stars in the LMC outskirts \citep{MagES3}. While our identification of a stellar overdensity is highly significant, the association of any individual object to the Magellanic system, as well as the exact distances/spectral types of individual objects are more uncertain. Spectroscopy of these faint and crowded objects requires JWST to further constrain their stellar types, distances, and dynamics, allowing for a more complete picture of this interesting system.
\vspace{0.5cm}

\noindent Acknowledgements: We thank Ian Smail for helpful suggestions regarding the source detection methods. We also thank Steven Willner and the referee for helpful suggestions that improved the submitted manuscript. JS acknowledges support from an undergraduate Arizona NASA Space Grant, Cooperative Agreement 80NSSC20M0041. RAW, SHC, and RAJ acknowledge support from NASA JWST Interdisciplinary Scientist grants NAG5-12460, NNX14AN10G and 80NSSC18K0200 from GSFC. CJC and NJA acknowledge support from the European Research Council (ERC) Advanced Investigator Grant EPOCHS (788113). MAM acknowledges the support of a National Research Council of Canada Plaskett Fellowship, and the Australian Research Council Centre of Excellence for All Sky Astrophysics in 3 Dimensions (ASTRO 3D), through project number CE17010001. CNAW acknowledges funding from the JWST/NIRCam contract NASS-0215 to the University of Arizona. We also acknowledge the indigenous peoples of Arizona, including the Akimel O'odham (Pima) and Pee Posh (Maricopa) Indian Communities, whose care and keeping of the land has enabled us to be at ASU's Tempe campus in the Salt River Valley, where much of our work was conducted. This work is based on observations made with the NASA/ESA/CSA James Webb Space Telescope. The data were obtained from the Mikulski Archive for Space Telescopes (MAST) at the Space Telescope Science Institute, which is operated by the Association of Universities for Research in Astronomy, Inc., under NASA contract NAS 5-03127 for JWST. These observations are associated with JWST programs 1176, 2736, and 2738. This work is based on observations taken by the RELICS Treasury Program (GO 14096) with the NASA/ESA HST, which is operated by the Association of Universities for Research in Astronomy, Inc., under NASA contract NAS5-26555. This work has made use of data from the European Space Agency (ESA) mission {\it Gaia} (\url{https://www.cosmos.esa.int/gaia}), processed by the {\it Gaia} Data Processing and Analysis Consortium (DPAC,
\url{https://www.cosmos.esa.int/web/gaia/dpac/consortium}). Funding for the DPAC has been provided by national institutions, in particular the institutions participating in the {\it Gaia} Multilateral Agreement. We also acknowledge the JWST ERO team responsible for providing these products.\footnote{https://www.stsci.edu/jwst/science-execution/approved-programs/webb-first-image-observations}

\software{Astropy \citep{astropy:2013, astropy:2018, astropy:2022}; EAZY \citep{eazy}; JWST Calibration Pipeline \citep{bushouse_howard_2022_7229890}; Photutils \citep{photutils}; ProFound \citep{robotham_profound}; Pypher \citep{pypher}; Pysynphot \citep{pysynphot}; SourceExtractor \citep{sextractor}; WebbPSF \citep{webbpsf}}

\facilities{Hubble and James Webb Space Telescope Mikulski Archive
\url{https://archive.stsci.edu}. All of the JWST data used in this paper can be found on MAST: \dataset[10.17909/b7hf-he53]{http://dx.doi.org/10.17909/b7hf-he53}. The RELICS HST data used in this paper can be found here: \dataset[10.17909/T9SP45]{http://dx.doi.org/10.17909/T9SP45}.}

\startlongtable
\begin{deluxetable*}{cccccccccccc} 
\label{candidates}
\tablecaption{Candidate LMC stars identified with star-galaxy separation and fit with IRTF spectral templates.}
\tablenum{1}
\tablehead{\multicolumn{3}{c}{} & \multicolumn{6}{c}{------------------------ PHOTOMETRY\tablenotemark{a} -----------------------} & \multicolumn{3}{c}{-------- PROPERTIES --------} \\ \colhead{id} & \colhead{RA} & \colhead{Dec} & \colhead{F090W} & \colhead{F150W} & \colhead{F200W} & \colhead{F277W} & \colhead{F356W} & \colhead{F444W} & \colhead{Type} & \colhead{$\chi^{2}_{\text{red}}$} & \colhead{$\mu$\tablenotemark{b}} \\ \colhead{} & \colhead{(deg)} & \colhead{(deg)} & \colhead{(mag)} & \colhead{(mag)} & \colhead{(mag)} & \colhead{(mag)} & \colhead{(mag)} & \colhead{(mag)} & \multicolumn{2}{c}{} & \colhead{(mag)}}
\startdata
1 & 110.86216700 & -73.47061220 & 24.77 & 24.33 & 24.53 & 25.24 & 25.49 & 25.82 & K7V & 0.66 & $18.1^{+0.3}_{-0.3}$\\
2 & 110.85557140 & -73.47091030 & 24.95 & 24.69 & 24.97 & 25.58 & 26.05 & 26.56 & K3V & 0.66 & $19.0^{+0.3}_{-0.3}$\\
3 & 110.71958260 & -73.49692810 & 25.05 & 24.82 & 25.03 & 25.67 & 26.16 & 26.50 & K3V & 0.48 & $19.1^{+0.3}_{-0.3}$\\
4 & 110.91536500 & -73.45760250 & 26.30 & 25.84 & 26.04 & 26.78 & 26.89 & 27.34 & K7V & 0.89 & $19.4^{+0.4}_{-1.2}$\\
5 & 110.67205510 & -73.50673290 & 26.63 & 26.24 & 26.26 & 26.70 & 27.31 & 27.31 & M1V & 0.81 & $18.7^{+0.5}_{-1.0}$\\
6 & 110.70759900 & -73.49949690 & 26.50 & 25.99 & 26.16 & 26.69 & 27.21 & 27.55 & M0V & 0.27 & $19.2^{+0.5}_{-0.7}$\\
7 & 110.67496690 & -73.50461890 & 25.96 & 25.68 & 25.89 & 26.54 & 26.99 & 27.46 & K3V & 0.19 & $19.9^{+0.3}_{-0.3}$\\
8 & 110.66512020 & -73.50600780 & 26.55 & 26.06 & 26.24 & 26.79 & 27.15 & 27.29 & M3V & 0.38 & $18.3^{+0.7}_{-0.7}$\\
9 & 110.70289650 & -73.49699630 & 25.90 & 25.62 & 25.84 & 26.37 & 26.86 & 27.11 & K3V & 0.33 & $19.8^{+0.3}_{-0.4}$\\
10 & 110.68002770 & -73.50094110 & 24.80 & 24.62 & 24.87 & 25.52 & 26.08 & 26.39 & K2V & 0.52 & $19.3^{+0.3}_{-0.3}$\\
11 & 110.89040420 & -73.45783110 & 26.02 & 25.52 & 25.64 & 26.25 & 26.63 & 27.08 & M0.5V & 0.30 & $18.6^{+0.4}_{-0.7}$\\
12 & 110.84920430 & -73.46677800 & 23.09 & 23.07 & 23.39 & 24.00 & 24.48 & 24.91 & G3Va & 0.18 & $18.1^{+1.5}_{-0.4}$\\
13 & 110.86553090 & -73.46272980 & 25.87 & 25.63 & 25.89 & 26.46 & 26.82 & 27.26 & K3V & 0.36 & $19.9^{+0.3}_{-0.3}$\\
14 & 110.72000470 & -73.49268410 & 25.59 & 25.15 & 25.32 & 25.84 & 26.24 & 26.62 & M0V & 0.24 & $18.2^{+0.6}_{-0.6}$\\
15 & 110.66805300 & -73.50092820 & 22.86 & 22.99 & 23.36 & 24.04 & 24.57 & 24.94 & G6.5V & 0.30 & $18.8^{+0.3}_{-0.5}$\\
16 & 110.72784760 & -73.48935170 & 26.17 & 25.77 & 25.96 & 26.51 & 26.95 & 27.29 & K7V & 0.06 & $19.5^{+0.3}_{-0.3}$\\
17 & 110.76571260 & -73.48077720 & 25.91 & 25.47 & 25.67 & 26.11 & 26.55 & 26.84 & M1V & 0.35 & $18.2^{+0.7}_{-0.9}$\\
18 & 110.87489520 & -73.45841050 & 25.19 & 25.08 & 25.36 & 25.99 & 26.40 & 26.88 & K2V & 0.03 & $19.8^{+0.3}_{-0.3}$\\
19 & 110.73000210 & -73.48620160 & 25.43 & 25.02 & 25.28 & 25.83 & 26.36 & 26.71 & K7V & 0.48 & $19.0^{+0.4}_{-0.3}$\\
20 & 110.71625750 & -73.48854100 & 26.05 & 25.60 & 25.79 & 26.36 & 26.69 & 26.99 & M0V & 0.47 & $18.6^{+0.7}_{-1.0}$\\
21 & 110.72057470 & -73.48737060 & 26.53 & 26.16 & 26.36 & 26.91 & 27.34 & 27.80 & K7V & 0.10 & $19.9^{+0.4}_{-0.3}$\\
22 & 110.80594110 & -73.46958100 & 25.72 & 25.34 & 25.58 & 26.18 & 26.53 & 26.90 & K7V & 0.43 & $19.2^{+0.5}_{-0.3}$\\
23 & 110.68128930 & -73.49252790 & 25.04 & 24.91 & 25.19 & 25.76 & 26.25 & 26.52 & K2V & 0.38 & $19.5^{+0.3}_{-0.3}$\\
24 & 110.80106310 & -73.46821900 & 24.89 & 24.63 & 24.82 & 25.40 & 25.87 & 26.26 & K3V & 0.09 & $18.8^{+0.3}_{-0.3}$\\
25 & 110.75536650 & -73.47725180 & 25.74 & 25.40 & 25.60 & 26.06 & 26.45 & 26.75 & K7V & 0.69 & $19.0^{+0.4}_{-0.8}$\\
26 & 110.68680600 & -73.49055430 & 26.78 & 26.26 & 26.46 & 26.97 & 27.35 & 28.14 & M0V & 0.82 & $19.8^{+0.4}_{-0.4}$\\
27 & 110.64580220 & -73.49817030 & 24.37 & 24.30 & 24.52 & 25.16 & 25.70 & 25.98 & K2V & 0.25 & $19.0^{+0.3}_{-0.3}$\\
28 & 110.82877260 & -73.46026270 & 22.95 & 23.07 & 23.44 & 24.08 & 24.55 & 24.97 & G1V & 0.17 & $18.9^{+0.3}_{-0.5}$\\
29 & 110.72096780 & -73.48176010 & 25.78 & 25.40 & 25.60 & 26.10 & 26.52 & 26.87 & K7V & 0.21 & $19.1^{+0.3}_{-0.3}$\\
30 & 110.67001300 & -73.49191860 & 26.37 & 25.94 & 26.09 & 26.62 & 27.05 & 27.44 & M0V & 0.12 & $19.3^{+0.4}_{-0.8}$\\
31 & 110.64265760 & -73.49445510 & 23.86 & 23.67 & 24.01 & 24.65 & 25.20 & 25.56 & K2V & 0.50 & $18.4^{+0.3}_{-0.3}$\\
32 & 110.81476720 & -73.45898890 & 23.29 & 23.33 & 23.65 & 24.26 & 24.72 & 25.14 & K0V & 0.13 & $18.5^{+1.4}_{-0.3}$\\
33 & 110.72384130 & -73.47836710 & 26.10 & 25.76 & 25.97 & 26.51 & 26.98 & 27.40 & K3V & 0.30 & $19.8^{+0.4}_{-0.4}$\\
34 & 110.63281260 & -73.49283500 & 25.68 & 25.16 & 25.40 & 25.96 & 26.50 & 26.94 & M0V & 0.67 & $18.9^{+0.3}_{-0.3}$\\
35 & 110.72055620 & -73.47503090 & 24.00 & 23.91 & 24.23 & 24.81 & 25.33 & 25.72 & K1V & 0.10 & $18.7^{+0.3}_{-0.3}$\\
36 & 110.66272890 & -73.48596090 & 25.92 & 25.65 & 25.85 & 26.44 & 26.96 & 27.41 & K5V & 0.06 & $19.8^{+0.3}_{-0.3}$\\
37 & 110.65067740 & -73.49085180 & 26.12 & 25.79 & 25.96 & 26.52 & 27.01 & 27.31 & K7V & 0.34 & $19.7^{+0.5}_{-0.3}$\\
38 & 110.84558040 & -73.44554930 & 24.30 & 24.28 & 24.61 & 25.22 & 25.65 & 26.09 & G3Va & 0.17 & $19.3^{+1.6}_{-0.4}$\\
39 & 110.80421890 & -73.45313020 & 24.13 & 24.09 & 24.44 & 24.98 & 25.48 & 25.86 & G3Va & 0.23 & $19.1^{+0.4}_{-0.4}$\\
40 & 110.67477770 & -73.47867250 & 25.45 & 25.14 & 25.37 & 25.95 & 26.45 & 26.82 & K3V & 0.08 & $19.4^{+0.3}_{-0.3}$\\
41 & 110.63467750 & -73.48880260 & 24.35 & 24.07 & 24.36 & 24.93 & 25.44 & 25.81 & K3V & 0.37 & $18.3^{+0.3}_{-0.3}$\\
42 & 110.79796380 & -73.45281710 & 26.56 & 25.94 & 26.08 & 26.73 & 27.12 & 27.56 & M1.5V & 0.33 & $19.1^{+0.3}_{-0.7}$\\
43 & 110.64595860 & -73.48242080 & 22.72 & 22.87 & 23.24 & 23.86 & 24.43 & 24.81 & G1V & 0.15 & $18.7^{+0.3}_{-0.5}$\\
44 & 110.83642120 & -73.44363830 & 25.95 & 25.57 & 25.77 & 26.27 & 26.58 & 27.00 & K7V & 0.47 & $19.1^{+0.4}_{-1.1}$\\
45 & 110.76016130 & -73.45909700 & 25.50 & 25.19 & 25.45 & 25.99 & 26.40 & 26.89 & K3V & 0.07 & $19.4^{+0.3}_{-0.4}$\\
46 & 110.72820980 & -73.46594990 & 24.27 & 24.38 & 24.58 & 25.10 & 25.70 & 26.12 & G3Va & 0.54 & $19.4^{+1.5}_{-0.3}$\\
47 & 110.78956710 & -73.45340770 & 26.71 & 26.31 & 26.46 & 27.05 & 27.36 & 27.61 & K7V & 0.58 & $19.2^{+0.8}_{-1.1}$\\
48 & 110.83748180 & -73.44288410 & 25.02 & 24.67 & 24.87 & 25.44 & 25.91 & 26.31 & K7V & 0.28 & $18.6^{+0.4}_{-0.3}$\\
49 & 110.65875080 & -73.47794840 & 26.18 & 25.71 & 25.84 & 26.43 & 26.88 & 27.12 & M1V & 0.36 & $18.7^{+0.6}_{-0.6}$\\
50 & 110.78230860 & -73.45339920 & 26.37 & 25.93 & 26.00 & 26.69 & 27.03 & 27.62 & M0V & 0.39 & $19.4^{+0.3}_{-0.3}$\\
51 & 110.78782340 & -73.45101940 & 23.53 & 23.51 & 23.82 & 24.41 & 24.94 & 25.31 & G3Va & 0.18 & $18.5^{+0.4}_{-0.4}$\\
52 & 110.85693660 & -73.43664670 & 26.00 & 25.70 & 25.91 & 26.42 & 26.84 & 27.40 & K3V & 0.20 & $19.9^{+0.3}_{-0.4}$\\
53 & 110.66720140 & -73.47507180 & 25.12 & 24.92 & 25.20 & 25.63 & 26.17 & 26.57 & K3V & 0.84 & $19.2^{+0.3}_{-0.4}$\\
54 & 110.81224460 & -73.44403790 & 23.88 & 23.77 & 24.05 & 24.59 & 25.04 & 25.46 & K2V & 0.26 & $18.4^{+0.3}_{-0.3}$\\
55 & 110.66261310 & -73.47706440 & 24.46 & 24.33 & 24.63 & 25.23 & 25.77 & 26.14 & K2V & 0.07 & $19.0^{+0.3}_{-0.3}$\\
56 & 110.78046660 & -73.45030070 & 26.49 & 26.12 & 26.31 & 26.82 & 27.17 & 27.61 & K7V & 0.25 & $19.8^{+0.3}_{-0.4}$\\
57 & 110.76385480 & -73.45338740 & 25.74 & 25.40 & 25.62 & 26.17 & 26.64 & 27.07 & K3V & 0.27 & $19.5^{+0.4}_{-0.4}$\\
58 & 110.79180220 & -73.44578380 & 23.77 & 23.71 & 24.06 & 24.67 & 25.13 & 25.57 & G3Va & 0.37 & $18.7^{+0.4}_{-0.4}$\\
59 & 110.77367240 & -73.44919120 & 25.65 & 25.25 & 25.47 & 26.12 & 26.56 & 27.03 & K7V & 0.67 & $19.3^{+0.4}_{-0.4}$\\
60 & 110.77943010 & -73.44784300 & 26.23 & 25.88 & 26.04 & 26.55 & 26.93 & 27.25 & K7V & 0.50 & $19.4^{+0.4}_{-0.8}$\\
61 & 110.62570660 & -73.47884900 & 22.68 & 22.76 & 23.14 & 23.70 & 24.24 & 24.64 & G2V & 0.13 & $18.2^{+0.6}_{-0.3}$\\
62 & 110.84937560 & -73.43260770 & 22.94 & 23.00 & 23.39 & 24.10 & 24.47 & 24.88 & G2V & 0.83 & $18.7^{+0.4}_{-0.5}$\\
63 & 110.75095580 & -73.45261390 & 25.81 & 25.45 & 25.65 & 26.19 & 26.60 & 27.07 & K7V & 0.17 & $19.2^{+0.4}_{-0.3}$\\
64 & 110.83870650 & -73.43454150 & 25.64 & 25.39 & 25.56 & 26.09 & 26.52 & 26.96 & K3V & 0.33 & $19.5^{+0.3}_{-0.4}$\\
65 & 110.67197890 & -73.46799340 & 26.70 & 26.22 & 26.38 & 26.86 & 27.31 & 27.90 & M0V & 0.36 & $19.6^{+0.4}_{-0.6}$\\
66 & 110.61303870 & -73.47875580 & 24.38 & 24.03 & 24.27 & 24.85 & 25.37 & 25.77 & K3V & 0.22 & $18.3^{+0.3}_{-0.4}$\\
67 & 110.79838860 & -73.43960650 & 26.59 & 26.07 & 26.24 & 26.82 & 27.23 & 27.61 & M0V & 0.37 & $19.1^{+0.5}_{-0.7}$\\
68 & 110.64934660 & -73.46977780 & 25.55 & 25.21 & 25.47 & 26.01 & 26.55 & 26.97 & K3V & 0.16 & $19.4^{+0.3}_{-0.3}$\\
69 & 110.81135150 & -73.43491940 & 25.24 & 24.98 & 25.16 & 25.76 & 26.18 & 26.65 & K3V & 0.11 & $19.2^{+0.3}_{-0.3}$\\
70 & 110.65634220 & -73.46544270 & 25.81 & 25.33 & 25.53 & 26.04 & 26.42 & 26.78 & M1V & 0.31 & $18.1^{+0.7}_{-0.8}$\\
71 & 110.61594180 & -73.47252690 & 25.59 & 25.32 & 25.55 & 26.07 & 26.55 & 26.83 & K3V & 0.30 & $19.5^{+0.3}_{-0.4}$\\
\enddata
\tablenotetext{a}{Reported photometry was measured as discussed in Section \ref{sec23} on PSF-matched NIRCam mosaics. Apparent magnitude uncertainties are not displayed because they are very small ($\lesssim$$0.01$ mag). All reported magnitudes use the AB system.}
\tablenotetext{b}{From the SED fits using templates from the IRTF Spectral Library \citep{spex2005, spex2009}. Distance moduli are calculated individually for each filter and averaged, where the lower uncertainty is the 16th percentile, the reported value is the 50th percentile, and the upper uncertainty is the 84th percentile of the cumulative probability distribution for a given star. The uncertainties are added in quadrature with an additional uncertainty of $0.25$ mag due to the SpeX flux calibration uncertainties. Only stars with $18 \leq \mu \leq 20$ mag are taken as candidates.}
\end{deluxetable*}

\bibliographystyle{aasjournal}
\bibliography{references_LMCstars}{}

\end{document}